\PassOptionsToPackage{unicode}{hyperref}
\PassOptionsToPackage{naturalnames}{hyperref}

\documentclass[twocolumn]{aastex7}
\usepackage{mathtools}
\usepackage{enumitem}
\usepackage{appendix}
\usepackage{xcolor}
\usepackage{subfigure}
\usepackage{graphicx}
\usepackage{amsmath}
\usepackage{figsize}
\usepackage{txfonts}
\usepackage[T1]{fontenc}
\usepackage{threeparttable}
\usepackage{CJKutf8}
\usepackage{caption}

\newcommand{\todo}{\ifmmode \text{\color{purple}\Huge{\(\bullet\)}} \else {\color{purple}{\Huge$\bullet$}}\fi}
\newcommand{\finish}{\ifmmode \text{\color{blue}\Huge{\(\bullet\)}} \else {\color{blue}{\Huge$\bullet$}}\fi}

\newcommand{\msun}{\mathrm{M_\odot}}

\newcommand{\tnr}[2]{{#1}_\mathrm{#2}}
\newcommand{\tnrd}[3]{{#1}_\mathrm{#2}^\mathrm{#3}}
\newcommand{\sfr}{\msun\ \rm yr^{-1}}
\newcommand{\lrp}[1]{\left({#1}\right)}
\newcommand{\robs}{\mathcal{R}_\mathrm{obs}}
\newcommand{\whz}{\mathrm{\ W\ Hz^{-1}}}
\newcommand{\sqdeg}{\mathrm{\ deg^2}}
\newcommand{\upda}[1]{\textcolor{black}{#1}}
\newcommand{\apjsa}[1]{\textcolor{black}{#1}}
\newcommand{\apjsb}[1]{\textcolor{black}{#1}}

\begin{document}

\begin{CJK*}{UTF8}{gkai}
\title{UNIONS Optical Identifications for VLASS Radio Sources in the Euclid Sky (UNVEIL) I. \\
A Catalog of $\sim146,000$ Radio Galaxies up to $z\sim5$}

\author[0009-0001-3910-2288]{Yuxing Zhong (仲宇星)}
\affiliation{Department of Physics, School of Advanced Science and Engineering, Faculty of Science and Engineering, Waseda University, 3-4-1, Okubo, Shinjuku, Tokyo 169-8555, Japan}
\email[show]{yuxing.zhong.astro@gmail.com}

\author[0000-0002-4377-903X]{Kohei Ichikawa}
\affiliation{Global Center for Science and Engineering, Faculty of Science and Engineering, Waseda University, 3-4-1, Okubo, Shinjuku, Tokyo 169-8555, Japan}
\affiliation{Department of Physics, School of Advanced Science and Engineering, Faculty of Science and Engineering, Waseda University, 3-4-1, Okubo, Shinjuku, Tokyo 169-8555, Japan}
\email[show]{ichikawa.waseda@gmail.com}

\author[0000-0002-9814-3338]{Hendrik Hildebrandt}
\affiliation{Ruhr University Bochum, Faculty of Physics and Astronomy, Astronomical Institute (AIRUB), German Centre for Cosmological Lensing, 44780 Bochum, Germany}
\email{hendrik@astro.ruhr-uni-bochum.de}

\author[0000-0001-8221-8406]{Stephen Gwyn}
\affiliation{National Research Council Herzberg Astronomy and Astrophysics, 5071 West Saanich Road, Victoria, B.C., V8Z6M7, Canada}
\email{Stephen.Gwyn@nrc-cnrc.gc.ca}

\author[0000-0002-7779-8677]{Akio K. Inoue}
\affiliation{Department of Physics, School of Advanced Science and Engineering, Faculty of Science and Engineering, Waseda University, 3-4-1, Okubo, Shinjuku, Tokyo 169-8555, Japan}
\affiliation{Waseda Research Institute for Science and Engineering, Faculty of Science and Engineering, Waseda University, 3-4-1,
Okubo, Shinjuku, Tokyo 169-8555, Japan}
\email{akinoue@aoni.waseda.jp}


\author[0000-0003-2984-6803]{Masafusa Onoue}
\affiliation{Waseda Institute for Advanced Study (WIAS), Waseda University, 1-21-1, Nishi-Waseda, Shinjuku, Tokyo 169-0051, Japan; Center for Data Science, Waseda University, 1-6-1, Nishi-Waseda, Shinjuku, Tokyo 169-0051, Japan}
\email{masafusa.onoue@aoni.waseda.jp}

\author[0000-0001-6186-8792]{Masatoshi Imanishi}
\affiliation{National Astronomical Observatory of Japan, Mitaka, Tokyo
181-8588, Japan.}
\affiliation{Department of Astronomical Science, Graduate University for
Advanced Studies (SOKENDAI), Mitaka, Tokyo 181-8588, Japan.}
\email{masa.imanishi@nao.ac.jp}

\author[0009-0001-9947-6732]{Taketo Yoshida}
\affiliation{Graduate School of Science and Engineering, Ehime University, 2-5 Bunkyo-cho, Matsuyama, Ehime 790-8577, Japan}
\email{tohru.nagao@gmail.com}

\author[0000-0001-9513-7138]{Martin Kilbinger}
\affiliation{Universit´e Paris-Saclay, Universit´e Paris Cit´e, CEA, CNRS, AIM, 91191, Gif-sur-Yvette, France}
\email{Martin.Kilbinger@cea.fr}

\author[0000-0001-5486-2747]{Thomas de Boer}  
\affiliation{Institute for Astronomy, University of Hawaii, 2680 Woodlawn Drive, Honolulu, HI 96822, USA}
\email{tdeboer@hawaii.edu}

\author[0000-0002-6639-6533]{Gregory S. H. Paek}  
\affiliation{Institute for Astronomy, University of Hawaii, 2680 Woodlawn Drive, Honolulu, HI 96822, USA}
\email{gpaek@hawaii.edu}

\author[0000-0001-5063-0340]{Yoshiki Matsuoka}
\affiliation{Research Center for Space and Cosmic Evolution, Ehime University, 2-5 Bunkyo-cho, Matsuyama, Ehime 790-8577, Japan}
\email{yk.matsuoka@cosmos.phys.sci.ehime-u.ac.jp}

\author[0000-0002-7402-5441]{Tohru Nagao}
\affiliation{Research Center for Space and Cosmic Evolution, Ehime
University, 2-5 Bunkyo-cho, Matsuyama, Ehime 790-8577, Japan}
\affiliation{Amanogawa Galaxy Astronomy Research Center, Kagoshima
University, 1-21-35 Korimoto, Kagoshima 890-0065, Japan}
\email{tohru.nagao@gmail.com}

\author[0000-0002-3531-7863]{Yoshiki Toba}
\affiliation{Department of Physical Sciences, Ritsumeikan University,
1-1-1 Noji-higashi, Kusatsu, Shiga 525-8577, Japan}
\affiliation{Academia Sinica Institute of Astronomy and Astrophysics,
11F of Astronomy-Mathematics Building, AS/NTU, No.1, Section 4,
Roosevelt Road, Taipei 10617, Taiwan}
\affiliation{Research Center for Space and Cosmic Evolution, Ehime
University, 2-5 Bunkyo-cho, Matsuyama, Ehime 790-8577, Japan}
\email{toba@fc.ritsumei.ac.jp}

\begin{abstract}
We present the results of optical identifications for Very Large Array Sky Survey (VLASS) radio sources employing the Ultraviolet Near Infrared Optical Northern Survey (UNIONS).
A cross-match between UNIONS and VLASS Epoch 2 catalogs yields $146,212$ radio galaxies down to $r=24.5$ mag over a wide area of $\sim4,200\sqdeg$ in the northern hemisphere.
We perform $g-$dropout selections and find $\gtrsim200$ sources at $z\sim4$ optimistically.
Of 63,019 sources with valid photometric redshifts, 8,692 are at $\tnr{z}{photo}\geq1$ and 1,171 are at $\tnr{z}{photo}\geq2$.
Based on spectral luminosities at 1.4 and 3~GHz using the valid UNIONS photo-\textit{z}, we identify $\sim49,000$ radio-loud AGNs (RLAGNs) with $\tnr{L}{1.4GHz}>10^{24}\whz$, and all radio galaxies at $z\geq1$ are RL.
Adopting radio loudness instead, 138,266 out of 146,212 UNIONS-VLASS radio galaxies are RL.
By virtue of the wide area and medium depth provided by UNIONS, our catalog greatly increases the number counts of RLAGNs at $z>1$.
We further cross-match the UNIONS-VLASS catalog with LOw-Frequency ARray Two-metre Sky Survey (LoTSS) at 144~MHz and VLA Faint Images of the Radio Sky at Twenty-cm (FIRST) at 1.4~GHz, yielding 101,671 UNIONS-VLASS-LoTSS, 79,638 -FIRST, and 64,672 -LoTSS-FIRST sources, respectively.
This multifrequency radio dataset reveals sources of various spectral shapes, including the steep spectrum of aged populations, the peaked spectrum of young populations, and the upturned spectrum that might be associated with transient sources.
The UNIONS-VLASS radio galaxy candidates will be covered by the Euclid wide survey, bringing about legacy values to benefit multi-faceted studies related to radio galaxies and beyond.
\end{abstract}



\section{Introduction} \label{sec:intro}
Supermassive black holes (SMBHs) are believed to ubiquitously reside in the center of most galaxies \citep{kormendy_coevolution_2013,mcconnell_revisiting_2013}, and a tight correlation between the mass of a SMBH and the properties of its host galaxy has been observed since the 1990s \citep{magorrian_demography_1998,ferrarese_fundamental_2000,gebhardt_relationship_2000,tremaine_slope_2002}.
The tight correlation implies that SMBHs grow in tandem with the parameters of their host galaxies, so-called ``co-evolution''.
How SMBHs co-evolve with their host galaxies and how they impact their host galaxy evolution are long-standing questions in modern astronomy. 

One key phenomenon to shape such co-evolution is called AGN feedback.
Such a feedback can lead to the outflowing and/or heating of the cold gas, suppressing the host galaxy star formation and self-regulating the growth of the SMBH \citep{fabian_observational_2012,alexander_what_2012,bischetti_suppression_2022,harrison_observational_2024,talbot_simulations_2024,husko_effects_2025}, although a compression of cold gas that enhances the star formation has also been observed \citep{shin_positive_2019,duncan_jwsts_2023}.
One mechanism of AGN feedback is through the coupling between the host galaxy ISM and the relativistic plasma, which is known as radio jets, launched from the central SMBH.
This radio/jet-mode feedback is considered the primary mode of AGN feedback in radio-loud active galactic nuclei (RLAGNs).

RLAGNs are often selected via the observed radio loudness $\robs(=\tnr{F}{radio}/\tnr{F}{optical}
>10)$ defined as the ratio of radio-to-optical flux densities \citep{kellermann_vla_1989,balokovic_disclosing_2012},
or via the 1.4~GHz spectral luminosity $\tnr{L}{1.4GHz}\gtrsim10^{24}\whz$; \citealt{peacock_statistics_1986,condon_14_1989,tadhunter_radio_2016,padovani_active_2017}).
In the local Universe, RLAGNs are found to preferentially reside in early-type galaxies correlated with high stellar masses, low star formation rates, and radiatively inefficient AGNs (\citealt{best_host_2005,heckman_coevolution_2014,gurkan_herschel-atlas_2015,pierce_agn_2022}; however, see also \citealt{ramos_almeida_are_2012,pierce_galaxy_2023}).
At higher redshifts ($z>1$), there is an increasing number of RLAGNs found to be radiatively efficient \citep[Eddington ratio $\tnr{\lambda}{Edd}\gtrsim3\times10^{-2}$;][]{merloni_synthesis_2008,ichikawa_wide_2021}
and their host galaxies can be star-forming or even starburst galaxies \citep{drouart_rapidly_2014,zhu_radio_2023,zhong_revisiting_2024}.
These high-redshift radio galaxies (HzRGs) at $z>1$ are found tightly correlated with galaxy mergers \citep{chiaberge_radio_2015,zhong_revisiting_2023,heckman_mergers_2024}, suggestive of merger-triggered star formation and launching of radio jets.
Therefore, RLAGNs across cosmic time, especially HzRGs, are valuable laboratories to study the interplay between the host galaxy evolution, SMBH growth, and radio jets, as well as the jet-accretion disc connections.

The rareness of RLAGNs ($10-20\%$ among all AGN populations; \citealt{kellermann_vla_1989,urry_unified_1995,ivezic_optical_2002,chiaberge_origin_2011}) requires large surveys to increase the sample size to statistically understand RLAGNs, their host galaxies, and the connections in between.
Over the past decades, following the groundbreaking Cambridge Catalogue of Radio Sources \citep[e.g.,][]{kellermann_spectra_1969}, many radio sky surveys have been conducted centered around $\sim1$ GHz, including the National Radio Astronomy Observatory (NRAO) Very Large Array (VLA) Sky Survey \citep[NVSS; 1.4~GHz;][]{condon_nrao_1998} and VLA Faint Images of the Radio Sky at Twenty-cm \citep[FIRST; 1.4~GHz;][]{becker_first_1995}, and Australian Square Kilometre Array Pathfinder (ASKAP) Evolutionary Map of the Universe \citep[EMU; 944 MHz;][]{norris_emu_2011} and Rapid ASKAP Continuum Survey \citep[RACS; $\sim900-1600$ MHz;][]{mcconnell_rapid_2020}.
The vast sky coverages ($\mathrm{>10,000\ deg^2}$) of these surveys offer great opportunities to build up an unbiased sample to elevate the statistical importance of studies related to RLAGNs.

Wide-area optical surveys are necessary to identify the optical counterparts of radio sources and determine their redshifts, either photometrically or spectroscopically.
Using the Sloan Digital Sky Survey (SDSS; \citealt{york_sloan_2000}), \citet{ivezic_optical_2002} and \citet{helfand_last_2015} identified the optical counterparts for the FIRST radio core sources at 1.4~GHz, adopting a positional cross-match between the two catalogs.
Due to the shallow optical sensitivity of SDSS $r$-band imaging ($r<22.2$ mag), the fraction of matches ($\tnr{f}{matches}$) is only $\sim30\%$, and a significant fraction of optically faint host galaxies may have been missed.
Using the Subaru Hyper Suprime-Cam (HSC) Subaru Strategic Program \citep[SSP;][]{aihara_hyper_2018}, \citet{yamashita_wide_2018} reaches $\tnr{f}{matches}=50.6\%$ optimistically for FIRST sources in the HSC field thanks to the depth of HSC reaching a limiting magnitude of 25.9 in \textit{i-}band.
Despite an improvement in associating radio sources with faint objects, elevating the fairness of the radio galaxy catalog, the survey area is $\mathrm{1,400\ deg^2}$ such that the catalog size is limited.

Other than the 1.4~GHz continuum, radio galaxy catalogs have been constructed centering at 3~GHz and 144~MHz.
Based on the largest sample of uniformly selected and spectrally confirmed quasars provided by SDSS, \citet{arnaudova_exploring_2024} identified 3,697 RL and
111,132 RQ sources in LOw-Frequency ARray (LOFAR) observations at 144~MHz via the radio loudness dichotomy.
With 384 h observations of the upgraded \textit{Karl G. Jansky} VLA, the VLA-COSMOS 3~GHz Large Project \citep{smolcic_vla-cosmos_2017} identifies 10,830 radio sources in the COSMOS field.
This project is equipped with a sensitivity better than that of FIRST by almost two orders of magnitude, greatly exploring the RQ populations.
These catalogs are advantageous explorations in the faint end of radio galaxies; however, they do not significantly expand the number counts of RLAGNs.

To improve both the sample size of RLAGNs and the unbiasedness of host galaxy populations, \apjsa{we search for UNIONS \citep{gwyn_unions_2025} optical counterparts for VLASS \citep{lacy_karl_2020} radio sources at 3~GHz.}
UNIONS reaches a depth of 24.7 mag in \textit{r-}band and covers an extremely wide area of $\sim5,000\sqdeg$ that fully overlaps with VLASS, bridging the gap between the shallow imaging of SDSS and the deep but slightly limited sky coverage of HSC-SSP.
VLASS is the state-of-the-art radio survey operated at 3~GHz in terms of spatial resolution, sensitivity, and total survey area.
The cross-match between these two catalogs will produce one of the largest RLAGN catalogs covering a wide redshift range.
Furthermore, both catalogs are ongoing projects, ensuring a great scalability of the UNIONS-VLASS catalog.

In this paper, we use radio galaxies for sources cross-matched between the UNIONS optical and VLASS radio catalogs, and define HzRGs as radio galaxies at $z>1$.
This paper is organized as follows.
In Section \ref{sec:data}, we introduce the optical and radio source catalogs used to build the radio galaxy catalog.
The cross-matches between the catalogs and the statistical results of our radio galaxy catalog are presented in Section \ref{sec:xmatched sources}.
We examine the optical and radio number counts of UNIONS-VLASS radio galaxies in Section \ref{sec:properties}, and we further present the selection of RLAGNs based on the observed radio loudness and spectral luminosity.
We show radio galaxy classifications using multifrequency radio datasets and discuss the high-\textit{z} sources based on the dropout technique in Section \ref{sec:discussion}.
A summary is presented in Section \ref{sec: summary}.
Throughout this paper, we assume a $\mathrm{\Lambda CDM}$ cosmology with $\Omega_m=0.309$, $\Omega_\Lambda=0.691$, and $H_0=\mathrm{67.7\ km\ s^{-1}\ Mpc^{-1}}$ \citep{planck_collaboration_planck_2016}.

\section{The Data} \label{sec:data}
\begin{figure*}[ht!]
\centering
\includegraphics[width=0.85\textwidth]{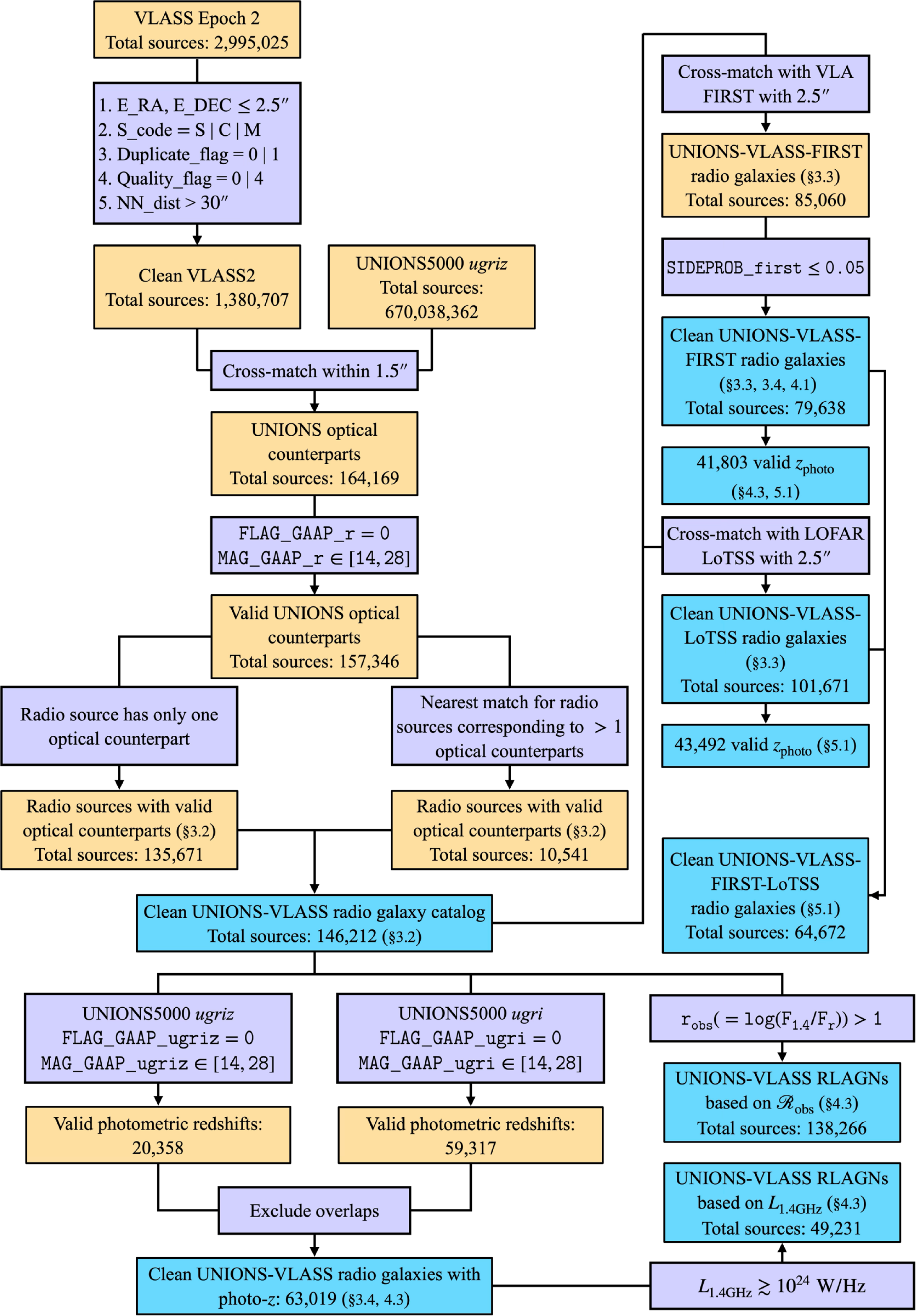}
\caption{Flowchart of the cross-match processes and sample selections (see \S\,\ref{sec:xmatched sources} for details).
In cyan boxes that describe the sample sizes, we supplement the sections where the corresponding sample is described and discussed.
\label{fig:flowchart}}
\end{figure*}

This exploration for radio galaxies is established on the basis of angular separation-based cross-match between the VLASS for radio sources at 3~GHz and UNIONS for optical counterparts.
The cross-matched UNIONS-VLASS catalog is set as our reference and further cross-matched with the FIRST 1.4~GHz and LoTSS 144~MHz radio source catalogs.
In Figure~\ref{fig:flowchart}, we summarize the flowchart that illustrates the criteria applied to each survey to filter out valid sources for the cross-match and to exclude invalid sources in the resultant cross-matches.

\subsection{UNIONS}
\label{sec: unions}
An elevation of the statistical importance of our radio galaxies catalog requires a wide ($>1000$~deg$^2$) and deep ($r>22$~mag) optical survey.
The Ultraviolet Near-Infrared Optical Northern Survey (UNIONS; \citealt{gwyn_unions_2025}), as a science collaboration between wide imaging surveys that cover the northern hemisphere, provides an excellent opportunity to abundantly discover the optical counterparts of radio sources.
The UNIONS integrates multiband (five broadbands, $u$, $g$, $r$, $i$, $z$) observations from Canada-France-Hawaii Telescope/Canada-France Imaging Survey (CFHT/CFIS, $u+r$), Subaru/Waterloo-Hawaii-IfA G-band Survey (WHIGS, \textit{g}), The Panoramic Survey Telescope \& Rapid Response System (Pan-STARRS, \textit{i}), and Subaru/Wide Imaging with Subaru HSC of the Euclid Sky (WISHES, \textit{z}).
The main UNIONS footprint is located at a declination (DEC.) $>30^\circ$ and
a galactic latitude of $|b|>25^\circ$, enclosing a total area of $4,861\sqdeg$.
Extended $u-$ and $z-$ (WISHES+) observations are being conducted to cover the sky area at $\mathrm{DEC.}<30^\circ$, elevating the sky coverage of UNIONS.
Albeit UNIONS is motivated by the \textit{Euclid} wide survey \citep{euclid_collaboration_euclid_2022}, efforts have been devoted to maintaining UNIONS as an independent survey with scientific values beyond what will be brought about by the \textit{Euclid}.

The $r$-band observation of CFHT/CFIS has an excellent median seeing of $0.69\arcsec$ and reaches a surface brightness limit of 28.4 $\mathrm{mag/arcsec^2}$ for extended sources.
It is set as the reference catalog by which the elliptical aperture is defined for the forced photometry applied to all five UNIONS bands.
Since the broadband observations of UNIONS are conducted under multiple telescopes, the Gaussian Aperture and Photometry ($\mathtt{GAaP}$; \citealt{kuijken_gravitational_2015}) is adopted to produce Gaussianized images for raw imaging data collected from observations at different instruments.
The forced photometry with Gaussian-weighted, elliptical apertures can then be properly used to perform homogeneous multiband photometric measurements simultaneously.
The depth of UNIONS is characterized by the $10\sigma$ point source depth measured through a $2.0\arcsec$ aperture.
The current median depth of UNIONS is $[u, g, r, i, \tnr{z}{PS}, \tnr{z}{HSC}] = [23.7, 24.5, 24.2, 23.8, 23.3, 23.3]$, to be compared with the original goal of $[u, g, r, i, z]=[23.6, 24.5, 23.9, 23.6, 23.4]$ \citep{gwyn_unions_2025}.

In this work, we use the latest internal 5-band data release UNIONS5000 (as of May 2025; $\mathrm{UNIONS5000}_{ugriz}$ hereafter) that includes $\sim20,600$ tiles with a total processed survey area of $\sim4,200\sqdeg$, covering both north (NGC) and south Galactic caps (SGC).
$\mathrm{UNIONS5000}_{ugriz}$ contains photometric measurements using forced photometry adopting an aperture size of $\mathrm{APER}=0.7\arcsec$ and $\mathrm{APER}=1.0\arcsec$, and an additional one with the optimal aperture $\mathrm{APER_{opt}}$.
In most cases, the optimal one is equal to the smallest possible aperture of $\mathrm{APER}=0.7\arcsec$.
In the case where the measurement adopting $\mathrm{APER}=0.7\arcsec$ fails, the optimal photometric measurement is then that using $\mathrm{APER}=1.0\arcsec$ (see also \citealt{kuijken_fourth_2019}).
We always use the optimal photometry for the data presented in this work.

\begin{figure}[ht!]
\centering
\includegraphics[width=\columnwidth]{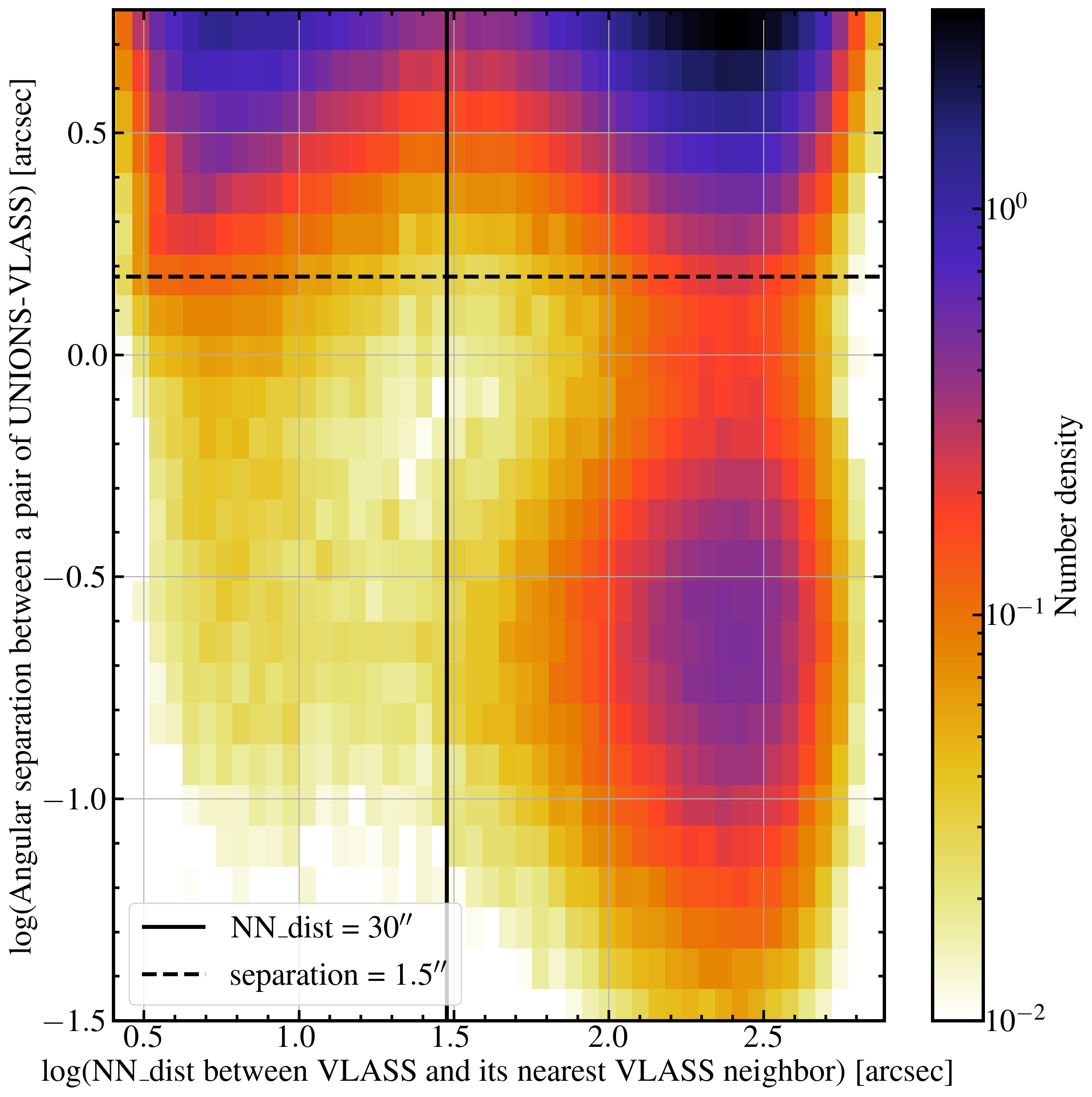}
\caption{The angular separation-$\mathtt{NN\_dist}$ plane for UNIONS-VLASS sources cross-matched with a search radius of $6''$.
The vertical lines indicate the $\mathtt{NN\_dist}$ cut of $30''$ and the horizontal line represents search radius of $1.5''$ to perform the cross-match (see \S\,\ref{subsec:completeness} and \ref{sec: optical-radio correspondence}).
\label{fig: nn_sep}}
\end{figure}

\subsection{VLASS}
\label{sec: vlass}
The VLASS (Very Large Array Sky Survey; \citealt{lacy_karl_2020}) observes the entire sky at a declination $>-40$ degrees, covering a total area of 33,885 $\mathrm{deg^2}$.
The observation frequency ranges $2-4$ GHz, 
with a central (reference) frequency of 3~GHz.
In baseline configurations in B and BnA to compensate for the projection effect, an angular resolution of $2.5\arcsec$ has been reached.
\apjsa{This wide-area survey is divided into three epochs, all of which were completed by October 2024, along with an additional half-epoch observation.\footnote{https://science.nrao.edu/vlass/library/vlass-epoch-4-science-case}
Combined with a single-epoch sensitivity (root mean square error, $1\sigma$) reaching 120~$\mu$Jy, these features make VLASS the widest and most sensitive radio sky survey with an excellent angular resolution at 3~GHz}
\apjsb{(\citealt{lacy_karl_2020}; see also Figure~1 in \citealt{de_gasperin_lofar_2021} for comparisons of VLASS with other radio surveys).}
Therefore, VLASS is ideal for serving as our reference radio source catalog.

VLASS data releases deliver quick look (QL) and single epoch (SE) images.
QL images are products of the rapid-CLEAN process with a cell size of $1\arcsec$.
SE images undergo deeper cleaning and self-calibration and reach a better cell size of $0.6\arcsec$.
However, since only an area of 1,000~$\mathrm{deg^2}$ is available in the SE images, we exclusively use QL products for the cross-match.
We use the epoch 2 catalog (VLASS2 hereafter) since astrometric errors and underestimates of flux densities were previously present in epoch 1, and the final release of epoch 3 is unsettled.

There are $\sim3$ M radio sources detected in VLASS2.
\apjsb{Prior to the cross-match, we adopt the following criteria to filter the VLASS2 catalog following the VLASS catalog user guide (see also \citealt{gordon_quick_2021})}
\begin{itemize}
    \item $\mathtt{E\_RA \& E\_DEC < 2.5''}$: astrometric uncertainties in both RA and DEC should be smaller than the angular resolution of $2.5\arcsec$;
    \item $\mathtt{Duplicate\_flag < 2}$: the source has no duplicated or has a preferred version of the duplicated counterpart within $2\arcsec$ search radius;
    \item $\mathtt{Quality\_flag == (0|4)}$: the source is unflagged or its peak flux density is higher than the integrated one ($\mathtt{Peak\_flux > Total\_flux}$);
    \item $\mathtt{S\_code != E}$: the source can be fitted with a single-Gaussian or multi-Gaussian components.
\end{itemize}
\apjsb{We further apply a cut on $\mathtt{NN\_dist}$ -- the angular separation between a VLASS source and its nearest neighbor in VLASS2 -- requiring it to be greater than $30''$ (see \S\,\ref{sec: nn_dist cut} for details).}
These criteria reduce the number of sources in ``clean'' VLASS2 to 1,380,707.
The VLASS2 catalog also includes the column $\mathtt{P\_sidelobe}$, which means the probability of a source being a sidelobe but misclassified as a real source.
As recommended in the VLASS Catalog User Guide, $\mathtt{Quality\_flag == 0}$ that indicates $\mathtt{P\_sidelobe < 0.1}$ is preferred, and the estimated contamination rate is 0.39\%.
In the cleanest criterion, $\mathtt{P\_sidelobe < 0.05}$ is applied, and the estimated contamination rate is 0.12\%.
For the final ``clean'' sample of 146,212 UNIONS-VLASS sources cross-matched (see \S\,\ref{sec: optical-radio correspondence} for details), adopting these two criteria results in a difference of 1,766 ($\sim1.2\%$) sources, which is statistically insignificant.
In the following of this work, without the loss of totality of our catalog, we present results adopting $\mathtt{P\_sidelobe < 0.1}$.

\subsection{FIRST}
\label{sec: first}
FIRST \citep{becker_first_1995} employs VLA to observe 8,444 $\mathrm{deg^2}$ square degrees of the NGC at 1.4~GHz and the survey area is limited to a declination of $<65$ degrees.
In the SGC, an area of $644\sqdeg$ has been observed, whereas the survey area does not overlap with the current internal data release of UNIONS5000 (see Figure~\ref{fig: sky coverage}).
In this work, we use the FIRST final data release catalog \citep{helfand_last_2015}.
In B-configuration, the survey reaches a spatial resolution represented by a circular Gaussian with full width at half maximum (FWHM) $5.4\arcsec$ with a pixel size of $1.8\arcsec$ over the northern sky.
For bright, pointlike sources with measured extents of $\leq2.0\arcsec$, the statistical astrometry is $0.22\arcsec$.
In a typical situation, the flux density detection limit is 1 mJy over the full footprint.

\subsection{LoTSS}
\label{sec: lotss}
To understand the evolution of radio galaxies, the spectral shape plays a crucial role, which requires multifrequency observations to construct the synchrotron spectral energy distribution (SED).
MHz-observations carried out by LOw-Frequency ARray are ideal supplements to our GHz-VLASS and FIRST data.
The LoTSS (LOFAR Two-metre Sky Survey; \citealt{shimwell_lofar_2017,shimwell_lofar_2022}) employs the LOFAR to observe 27\%  of the northern sky at $120-168$ MHz, with a central (reference) frequency of 144~MHz.
The sky coverage has two separate regions that span 4,178 and 1,457, $\mathrm{deg^2}$ respectively.
The full bandwidth Stokes I continuum imaging of LoTSS yields images at a spatial resolution of $6.0\arcsec$, centered at a frequency of 144~MHz.
The median detection sensitivity of the survey is 89 $\mathrm{\mu Jy\ beam^{-1}}$ with a typical positional uncertainty smaller than $0.2\arcsec$.
In this work, we use the latest and widest coverage catalog, LoTSS DR2 \citep{shimwell_lofar_2022}, which contains 4,396,228 radio sources derived from integrated intensity maps in Stokes I polarization.

\section{UNIONS-VLASS Radio Galaxy Catalog} \label{sec:xmatched sources}
In this section, we describe how we construct the UNIONS-VLASS radio galaxy catalog and how we supplement this catalog with multifrequency radio surveys, including FIRST and LoTSS.
We also describe the key information to be presented in our catalog, including redshifts, luminosities, spectral indices, and how a radio source is classified as an extended source.



\begin{figure*}[tp!]
\includegraphics[width=\textwidth]{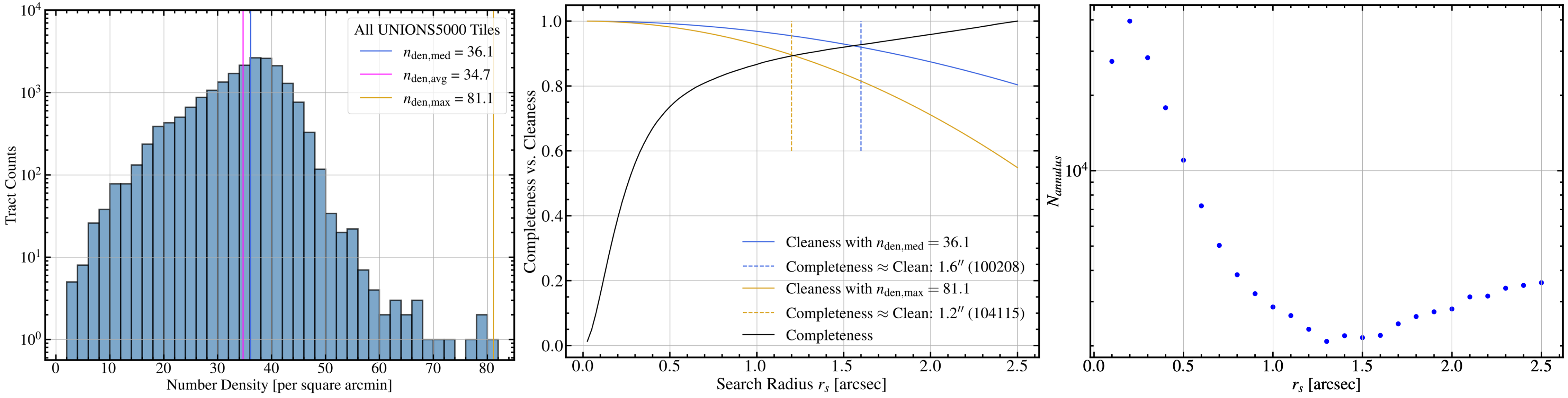}
\caption{\textit{Leftmost}: Tile-wise number densities $\tnr{n}{den}$ of galaxies per $\mathrm{arcmin^2}$ for UNIONS tiles in the North Galactic Cap.
The violet, blue, and orange lines indicate average, median, and maximal $\tnr{n}{den}$, respectively.
\textit{Middle}: completeness and cleanness as a function of the search radius $\tnr{r}{s}$.
The vertical dashed lines mark where the cleanness approximately intersects the completeness, indicating a trade-off between the totality of the catalog and the exclusion of the contamination.
\textit{Rightmost}: differential distribution of separation between UNIONS optical and VLASS radio sources including radio sources that are associated with multiple optical counterparts.
\label{fig: completeness}}
\end{figure*}

\subsection{$\mathtt{NN\_dist}$ Cut}
\label{sec: nn_dist cut}
\apjsa{
Radio galaxies with low radio luminosities often have their jets confined by the ISM and rapidly disrupted \citep{fanaroff_morphology_1974,turner_energetics_2015}.
Resultantly, the position of the peak surface brightness can be close to the host galaxy.
The nearest neighbor cross-match is thus a straightforward and efficient approach for identifying the optical counterpart of a radio source \citep{williams_lofar_2019}.
On the other hand, powerful radio jets can quickly overcome the confinement imposed by the ISM of their host galaxies without experiencing significant disruption \citep{mukherjee_relativistic_2016}.
Consequently, they often propagate far from their host galaxies and terminate in strong shocks -- referred to as radio hotspots -- where the observed radio surface brightness typically peaks \citep{turner_dynamics_2023}.
Identifying the optical counterparts of such radio sources necessitates a larger search radius, which, in turn, inevitably brings about more contaminants \citep{banfield_radio_2015,gordon_quick_2023}.}

\apjsa{
To statistically minimize chance overlaps between optical sources and radio sources exhibiting large-scale, complex structures, we first perform a cross-match between UNIONS and VLASS2 catalogs adopting the criteria present in \citet{gordon_quick_2021} and choosing a search radius of $6''$.
This cross-match yields $\sim550,000$ sources.
The angular separation between a pair of UNIONS-VLASS sources versus $\mathtt{NN\_dist}$ of the corresponding VLASS source is shown in Figure~\ref{fig: nn_sep}.
It is clear that the number of sources that are cross-matched shows a bimodal distribution along $\mathtt{NN\_dist}$. 
Cross-matched UNIONS-VLASS pairs with $\mathtt{NN\_dist\leq30''}$ tend to have large angular separations between their UNIONS and VLASS centroids.
We randomly select 36 VLASS sources with $\mathtt{NN\_dist<10''}$ and show their QL images, overlaid with VLASS and UNIONS source centroids, in Figure~\ref{fig: vlass nn imaging}. 
These radio sources often exhibit complex and extended morphology, such as core-jet and double-jet structures. 
VLASS could capture the centroid of the lobe or radio hotspot, naturally leading to the large offsets between UNIONS and VLASS.
We, hence, choose a $\mathtt{NN\_dist}$ cut of $30''$ to avoid false positive associations between UNIONS-VLASS caused by evolved radio structures that can extend well beyond the host galaxy scale and overlap with unrelated galaxies.}

\apjsb{It should be noted that this cut can remove a large fraction of bright, evolved radio galaxies at low-$z$.
Quantitatively, for a given search radius, let \textit{M} denote the number of cross-matched sources with $\mathtt{NN\_dist\leq30''}$ and \textit{N} the number with $\mathtt{NN\_dist>30''}$.
The fractional loss of the completeness can then be expressed as $\tnr{f}{loss}=M/(N+M)$.
As the search radius decreases from $6''$ to $1.5''$, $\tnr{f}{loss}$ decreases from $\sim25\%$ -- roughly the fraction of sources with $\mathtt{NN\_dist\leq30''}$ in the VLASS2 catalog -- to approximately $12\%$, while the number drops from 64,275 to 19,539.
Therefore, in addition to $\mathtt{NN\_dist}$, the elimination of extended radio sources with complex structures is also sensitive to the search radius resulting from the design of the cross-match methodology described in the following sections.
Consequently, we see that UNIONS-VLASS RGs are absent from the bright end ($\tnr{F}{3GHz}>3000$ mJy; Figure~\ref{fig: radio number counts}) and could be biased toward compact, young radio AGN populations (Figures~\ref{fig: alpha_lv}--\ref{fig: alpha_fv}; see \S\,\ref{sec:caveat} for details).}

\subsection{Contamination and Completeness} \label{subsec:completeness}
Since the galaxies are randomly distributed throughout the whole sky, the one-to-one correspondence between the radio and optical sources could be a false positive.
To statistically circumvent such contamination, we estimate the cleanness and completeness of the catalog in the following way.

We first estimate the number density of galaxies in each UNIONS tile lying within the VLASS field.
This gives a maximum number density of $\tnr{n}{den,max}=81.1$ per $\mathrm{arcmin^2}$ and a median value of $\tnr{n}{den,med}=36.1$ per $\mathrm{arcmin^2}$ (leftmost panel of Figure~\ref{fig: completeness}).
We perform an angular separation-based cross-match, choosing a search radius of $2.5\arcsec$.
We remove optical counterparts without valid photometry in any of \textit{ugriz} and multiple optical counterparts corresponding to a single radio source.
The total number of clean UNIONS-VLASS radio-optical pairs is then $\tnr{N}{tot}$, and the separation between each pair is defined as $\tnr{r}{sep}$.
Within the search radius of $r_{s}$, the number of sources with $\tnr{r}{sep}\leq\tnr{r}{s}$ is $\tnr{N}{s}$, and these sources occupy a total area of $\tnr{A}{s}=\pi\tnr{r}{s}^2\tnr{N}{s}$ $\mathrm{arcsec^2}$.
The completeness of UNIONS-VLASS is then represented by $\tnr{N}{s}/\tnr{N}{tot}$ and the maximum is 100\%.
The number of optical sources adjacent to radio sources by coincidence is $\tnr{N}{rand}=\tnr{A}{s}\tnr{n}{den}$.
The cleanness is then calculated by $(\tnr{N}{tot}-\tnr{N}{rand})/\tnr{N}{tot}$.
The $\tnr{r}{s}$ at which the cleanness intersects the completeness represents a compromise between the totality of the catalog and the contamination control (middle panel of Figure~\ref{fig: completeness}).
With $\tnr{n}{den,med}=36.1$ per $\mathrm{arcmin^2}$, the contaminated cross-matches arising from chance coincidence is 7.5\%, approaching an estimated completeness of 92.7\% at $\tnr{r}{s}\sim1.6\arcsec$.
When $\tnr{n}{den,max}=81.1$ per $\mathrm{arcmin^2}$ is adopted, $\tnr{r}{s}\sim1.2\arcsec$ is preferred, and the contamination rate is 9.1\% while the totality of the catalog is 89.3\%.

Without excluding optical counterparts associated with a single radio source, we show the differential distribution of the separation between UNIONS and VLASS sources (rightmost panel of Figure~\ref{fig: completeness}).
At $\tnr{r}{s}>1.5\arcsec$, the increase in the search radius is accompanied by an increase in the number of optical counterpart counts in the annulus, suggesting that these optical counterparts are likely to be coincident associations rather than real matches.
The threshold of $\tnr{r}{s}\sim1.5\arcsec$ is consistent with the completeness-cleanness trade-off of $1.6\arcsec$ adopting a median galaxy number density of $36.1$ per $\mathrm{arcmin^2}$.
\apjsa{This value is comparable to the findings of \citet{gordon_quick_2023}, who reported that the probability that an AllWISE source is the true postive counterpart to an unresolved VLASS radio source reaches 80\% at a positional uncertainty of $1.8''$.}

\begin{table}[tp!]
\apjsa{
\caption{Summary of the cross-match results}
\label{tab: cross-match}
\begin{center}
\begin{tabular}{ccccc}
\hline 
$\tnr{N}{VLASS2}$ & $\tnr{N}{UNIONS}$ & $\tnr{N}{matches}$ & $\tnr{f}{matches}$ & Area \\
 & & & & [$\mathrm{deg^2}$] \\
\hline
246,325 & 670,038,362 & 146,212 & 0.594 & $\sim4,200$ \\
\hline
\smallskip
\end{tabular}
\end{center}
{From left to right, the column represents number of VLASS2 sources, number of UNIONS sources, number of matches, fraction of matches, and the total cross-matched area.}}
\end{table}

\subsection{Clean UNIONS-VLASS Radio Galaxies}
\label{sec: optical-radio correspondence}
\apjsa{Choosing an angular separation of $\Delta\theta=\sqrt{(\Delta\alpha\cos\delta)^2+(\Delta\delta)^2}\leq1.5\arcsec$ as the search radius, our cross-match between the UNIONS and VLASS Epoch 2 catalogs immediately yields 164,169 optical counterparts of radio sources.}
This search radius represents a trade-off between the completeness of the catalog and a statistical exclusion of optical sources coincidentally associated with radio sources, which is described in \S\,\ref{subsec:completeness}.

To further make a clean radio galaxy catalog, we first eliminate optical sources without valid photometry in $r$-band since the UNIONS $\mathtt{GAaP}$ catalog is constructed upon the successful CFHT $r$-band detections.
This reduces the number of optical sources to 157,346.
Out of these sources, there are 135,671 UNIONS sources uniquely associated with VLASS sources, and they are flagged with $\mathtt{0}$ in the column $\mathtt{flag\_multi\_opt}$.

If a radio source has multiple optical counterpart candidates, we select the nearest one as the most likely true match.
This yields 10,541 radio sources with multiple optical counterparts within a search radius of $1.5\arcsec$, which are flagged with a value of $\mathtt{1}$ in the column $\mathtt{flag\_multi\_opt}$.
These multiple optical counterparts are also considered merging or interacting galaxy candidates, and they will be investigated in detail in future work.
\apjsa{Given the general compactness of our UNIONS-VLASS RGs (see \S\,\ref{sec: radio source counts} and \S\,\ref{subsubsec:radio color-color}), this nearest match method is expected to be valid in most cases, as jets and jet-induced shocks have not propagated far from their host galaxies.
However, this methodology does not fully rule out coincidental associations with unrelated optical counterparts, particularly in cases where RGs have developed jet-lobe structures that can lead to large offsets between radio-emitting regions and host galaxies.
As a result, the centroid of the radio emission captured by VLASS may be the radio hotspot and appears to be closer to an unassociated galaxy (up to approximately $6\%$ based on $\sim1,000$ \textit{Euclid}-VLASS RGs; Zhong et al. in prep.).}

The results of the cross-match are summarized in Table~\ref{tab: cross-match}.
Our angular separation-based cross-match generated 146,212 VLASS radio sources matched to UNIONS optical counterparts.
Defining the fraction of matched VLASS sources in the area that overlaps with the UNIONS as the fraction of matches $(\tnr{f}{matches})$, the corresponding value is 0.594, which is the maximum to be considered.
By virtue of the higher sensitivity and wide sky coverage provided by the UNIONS, this rate is higher than those in the previous studies cross-matching FIRST with SDSS ($\sim30\%$; \citealt{ivezic_optical_2002,helfand_last_2015}).
This fraction of matches also exceeds the cross-matching FIRST with Subaru HSC ($\sim47\%$; \citealt{yamashita_wide_2018}), even HSC optical number counts share a similar trend as the UNIONS (see \ref{subsec: optical source counts} for further discussions).
This could be explained by a combination of several factors, including a higher sensitivity of VLASS over FIRST and a larger search radius ($1.5\arcsec$) of UNIONS-VLASS than $1\arcsec$ of HSC-FIRST.

In summary, our cross-match between UNIONS and VLASS catalogs yields 146,212 radio galaxies. 
This ``clean'' UNIONS-VLASS radio galaxy catalog is the final sample to be presented in this work.

\begin{figure*}[htp!]
\includegraphics[width=\textwidth]{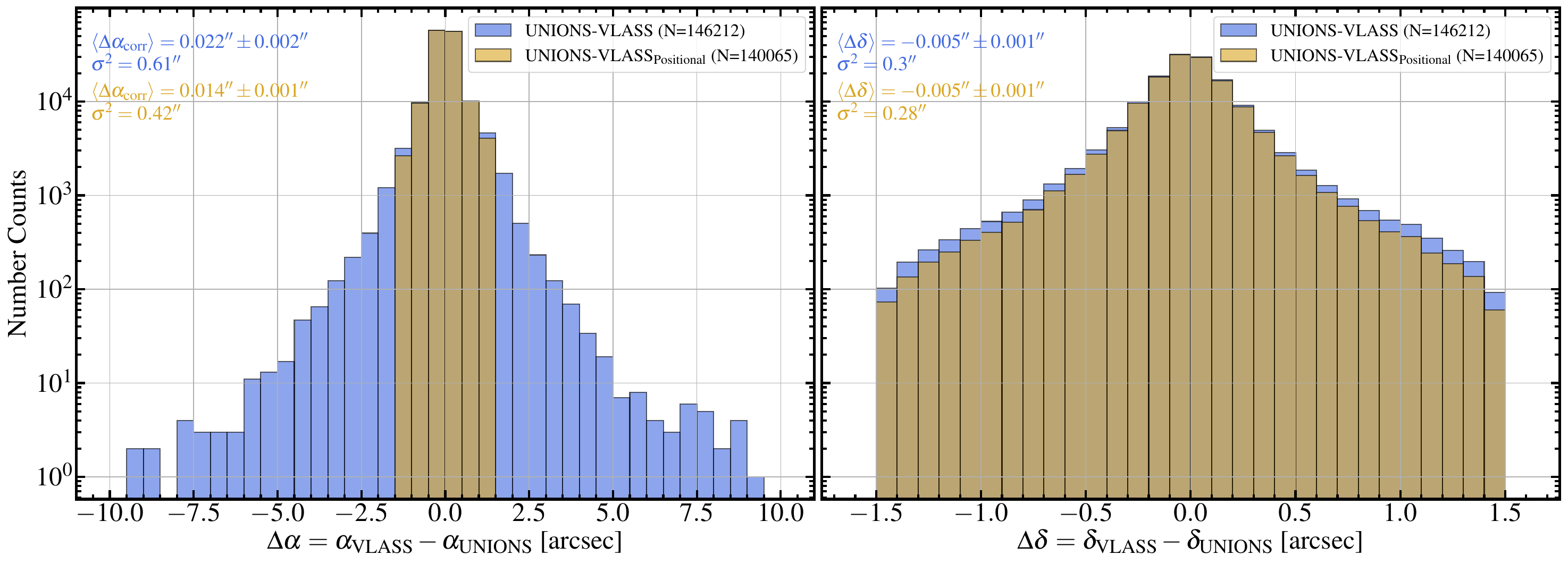}
\caption{Astrometric offsets in R.A. ($\Delta\alpha$) and DEC. ($\Delta\delta$) for 146,212 ``clean'' UNIONS-VLASS sources (blue) associated via $\sqrt{\Delta\alpha^2+\Delta\delta^2}\leq1.5\arcsec$ and subsample of 140,065 source adopting positional cross-match (orange; see texts for details).
Mean values of the offsets are embedded in the bracket and $\sigma$ represents the standard deviation.
The typical astrometric offset of VLASS Epoch 1 is $\sim0.5\arcsec$ at $\mathrm{DEC.>-20\ deg^2}$.
\label{fig: astrometry}}
\end{figure*}

\subsection{UNIONS-VLASS Astrometry}
\label{subsec:astrometry}
This vast radio galaxy catalog offers an excellent opportunity to investigate the systematic astrometric offsets between UNIONS and VLASS catalogs.
The offsets along R.A. and DEC. are residuals of a VLASS position minus a UNIONS position: $\Delta\alpha=\mathrm{\alpha_{VLASS}-\alpha_{UNIONS}}$ and $\Delta\delta=\mathrm{\delta_{VLASS}-\delta_{UNIONS}}$.
However, our cross-match method selects associated pairs based on the angular separation ($\Delta\theta$) between radio and optical sources.
Because the positional difference along the R.A. on the celestial sphere has a declination dependence, this calculation has to be corrected for the projection effect and is given by $\Delta\theta=\sqrt{(\tnr{\Delta\alpha}{corr})^2+(\Delta\delta)^2}$, where $\tnr{\Delta\alpha}{corr}=(\mathrm{\alpha_{VLASS}-\alpha_{UNIONS}})\times\cos(\tnr{\delta}{UNIONS})$.
Therefore, for $\Delta\alpha=\mathrm{\alpha_{VLASS}-\alpha_{UNIONS}}$, the value can exceed the search radius of $1.5\arcsec$, and such an excess becomes more severe for sources at higher declinations, as shown in the left panel of Figure~\ref{fig: astrometry app}.

To make a fair comparison of the astrometric accuracy in R.A. with other catalogs adopting positional cross-match that simply calculates the residuals between coordinates, we then make a subsample of ``clean'' UNIONS-VLASS sources.
This subsample contains 140,065 UNIONS-VLASS sources that have their optical counterparts lying within a circle of a projected radius $1.5\arcsec$ on the sky and centered on the radio source.  
This reduces the catalog size by $\sim4.2\%$, which is statistically insignificant.
The distributions of $\Delta\alpha$ and $\Delta\delta$ of these sources are shown by orange histograms in Figure~\ref{fig: astrometry}.

The standard deviations of 140,065 UNIONS-VLASS sources whose $\sqrt{\Delta\alpha^2+\Delta\delta^2}\leq1.5\arcsec$ are $0.44\arcsec$ in R.A. and $0.29\arcsec$ in DEC.
On the other hand, for 146,212 ``clean'' UNIONS-VLASS sources, the offset along the R.A. retards to $0.61\arcsec$ while the change in the offset along DEC. is negligible.
These offsets are much worse than the positional uncertainties (20 mas) of the CFHT $u-$ and $r$-band imaging surveys employed by UNIONS \citep{gwyn_unions_2025}.
For the VLASS catalog, by comparing the Epoch 1 catalog with Gaia DR2, the VLASS component positional accuracy is $\sim0.5\arcsec$ at $\mathrm{DEC.>-20\ deg^2}$.
Therefore, the current standard deviations of the positional errors for our UNIONS-VLASS sources are reasonable compared to the entire VLASS survey, suggesting that the astrometric offsets between optical and radio pairs are attributed to the intrinsic uncertainties of VLASS QL images.

In VLASS Epoch 1, the positional error peaks at $0.0\arcsec$ in R.A. and peaks at $-0.25\arcsec$ in DEC.
Our UNIONS-VLASS sources based on Epoch 2 observations show a similar trend in R.A. with an improved astrometric accuracy in DEC. ($\tnr{\Delta\delta}{peak}\sim0.0\arcsec$) compared to VLASS Epoch 1.
However, there is no significant improvement in the typical astrometry of $\sim0.5\arcsec$, which is much worse than that of FIRST ($0.2\arcsec$).
This is more of a problem of the rapid CLEAN procedure adopted by VLASS QL products.
More accurate astrometry can be expected when Epoch 3 and epoch-concatenated data are delivered, as well as for the SE images.

\subsection{Cross-matches between Surveys}
Once the unique optical counterpart catalog of VLASS sources (146,212 ``UNIONS-VLASS Radio Galaxies'' in Figure~\ref{fig:flowchart}) is constructed, we then perform additional cross-match with the VLA/FIRST 1.4~GHz survey and LOFAR/LoTSS 144~MHz survey.
We search for FIRST 1.4~GHz radio counterparts around our UNIONS-VLASS radio galaxies by requiring the angular separation between a pair of VLASS and FIRST sources to be within a $2.5\arcsec$ radius. 
This results in a total of 85,060 matched sources.
We then remove the sources containing possible sidelobes, using the sidelobe probability $\mathtt{P(S)}$ presented in the FIRST catalog.
We choose $\mathtt{0.00 < P(S) < 0.05}$ as the threshold, by which 76.3\% of sources are considered true positives, and this results in 79,638 radio sources identified in UNIONS, VLASS, and FIRST.

We search for LoTSS 144~MHz radio counterparts for the UNIONS-VLASS radio galaxies by applying a positional cross-match of $<2.5\arcsec$ radius, consistent with the VLASS beam size.
This cross-match results in 101,671 radio galaxies identified in UNIONS, VLASS, and LoTSS.
Finally, there are 64,672 UNIONS-VLASS-FIRST-LoTSS radio galaxies.

\subsection{Catalog Information}

\begin{figure*}[tp!]
\centering
\includegraphics[width=0.8\textwidth]{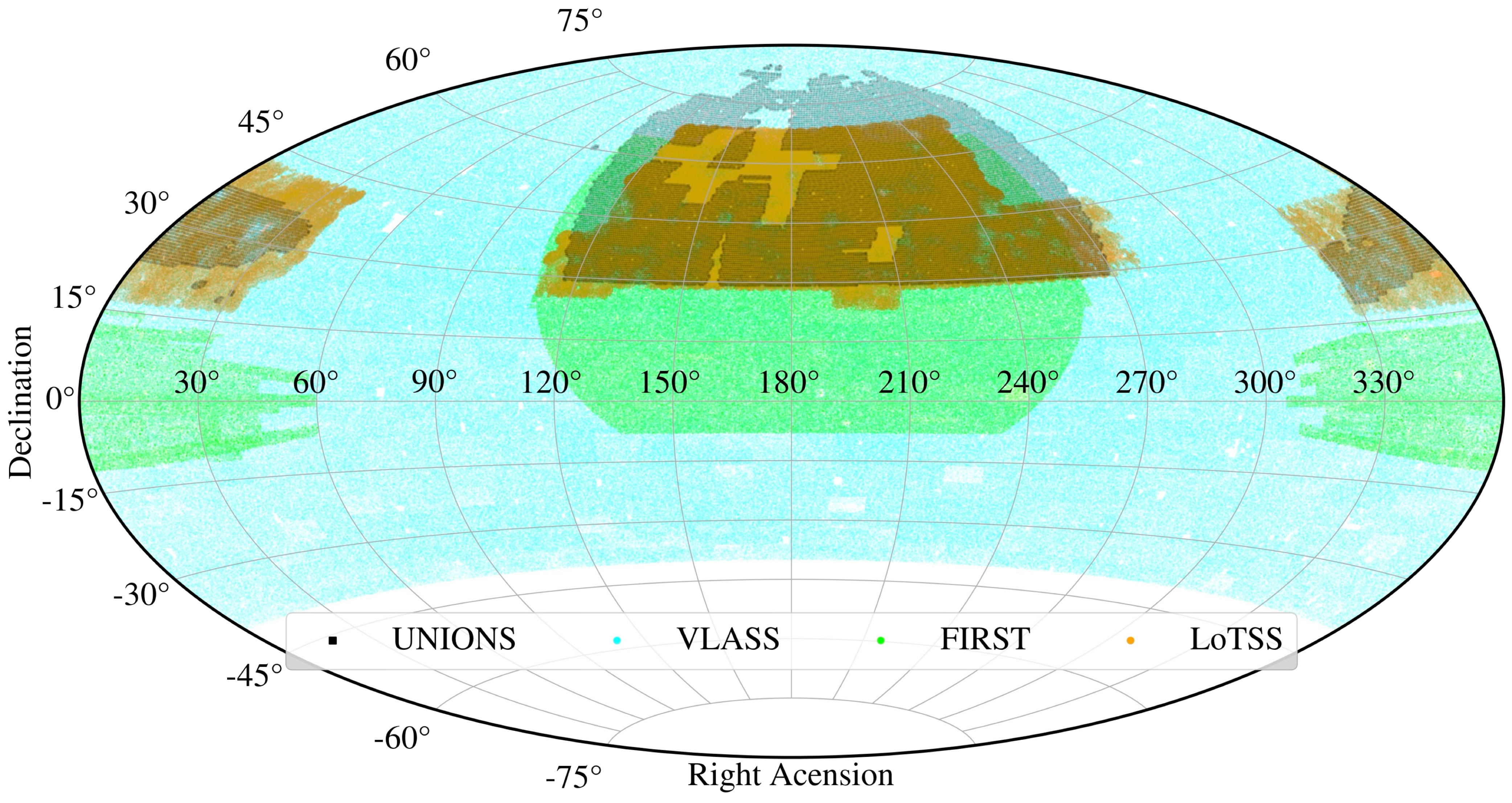}
\caption{The sky coverages of UNIONS (black), VLASS (cyan), FIRST (green), and LoTSS (orange).
\label{fig: sky coverage}}
\end{figure*}

In Figure~\ref{fig: sky coverage}, we show the sky coverages of UNIONS (black), VLASS (cyan), FIRST (green), and LoTSS (orange).
The presented ``clean'' UNIONS-VLASS catalog contains 146,212 radio galaxies that are distributed amongst an area of $\sim4,200\sqdeg$, reaching a number density of $\sim34\ \mathrm{deg^{-2}}$.
In the following subsections, we introduce the key information -- as well as how they are computed -- to be included in the catalog.

\subsubsection{Valid Photometric Redshifts of UNIONS}
The photometric redshifts presented in UNIONS are estimated from Bayesian Photometric Redshift \citep[BPZ;][]{benitez_bayesian_2000}.
The source with reliable photometric redshift should have valid photometric measurements in all five bands ($ugriz$): (1) the photometry is not flagged (column $\mathtt{FLAG\_GAaP\_{ugriz} == 0}$) and (2) the AB magnitudes within the optimal Gaussianized aperture should be valid measurements (column $\mathtt{MAG\_GAaP\_{ugriz} != (-99|99})$), where $\mathtt{99}$ indicates that the source is below the detection limit and $\mathtt{-99}$ stands for an absence of the observational data.
Therefore, of 146,212 ``clean'' UNIONS-VLASS sources, 20,358 have photo-$z$ with good reliabilities.
These sources are labeled with $\mathtt{0}$ in the column $\mathtt{flag\_z\_photo}$.

To maximize the availability of the redshift information in our catalog, we further make use of the 4-band version of UNIONS5000 ($\mathrm{UNIONS5000}_{ugri}$ hereafter) in which the photometric redshifts are estimated based on $ugri$, without the $z-$band photometry.
We perform an angular separation-based cross-match between VLASS2 and $\mathrm{UNIONS5000}_{ugri}$ and exclude the cases where multiple optical sources are associated with one radio source.
This is to circumvent erroneously associating the sources in $\mathrm{UNIONS5000}_{ugri}$ with another one in $\mathrm{UNIONS5000}_{ugriz}$.
This cross-matching yields $\upda{118,370}$ radio galaxies whose radio sources have a unique optical counterpart within the search radius.
Adopting the criteria $\mathtt{FLAG\_GAaP\_{ugri} == 0}$ and $\mathtt{MAG\_GAaP\_{ugri} != (-99|99})$, $\upda{59,317}$ have valid redshifts.
Excluding those with valid photo-$z$ in $\mathrm{UNIONS5000}_{ugriz}$ as well, this adds up $\upda{42,661}$ photo-$z$ estimates to our catalog (indicated by $\mathtt{flag\_z\_photo==1}$).
And in total, and there are $\upda{63,019}$ reliable $\tnr{z}{photo}$.

\subsubsection{Photometric versus Spectroscopic Redshifts} \label{subsec:photo-z}
\citet{gwyn_unions_2025} compared $\tnr{z}{photo}$ estimates with the spectroscopic redshifts collected from publicly available datasets in the UNIONS footprint.
The bias and scatter of photo-$z$ are represented by the median and normalized median absolute deviation (NMAD) of the quantity $\frac{\Delta z}{1+z}=\frac{\tnr{z}{spec}-\tnr{z}{photo}}{1+\tnr{z}{spec}}$, respectively.
Correspondingly, the bias is $-0.008$ and the scatter is $0.061$ for UNIONS sources with $\tnr{z}{photo}\lesssim1.6$.
If selecting outliers by $\frac{|\Delta z|}{1+z}>0.15$, then $16.3\%$ of the sources are considered having no reliable photo-$z$ estimates.

\begin{figure}[tp!]
\includegraphics[width=\columnwidth]{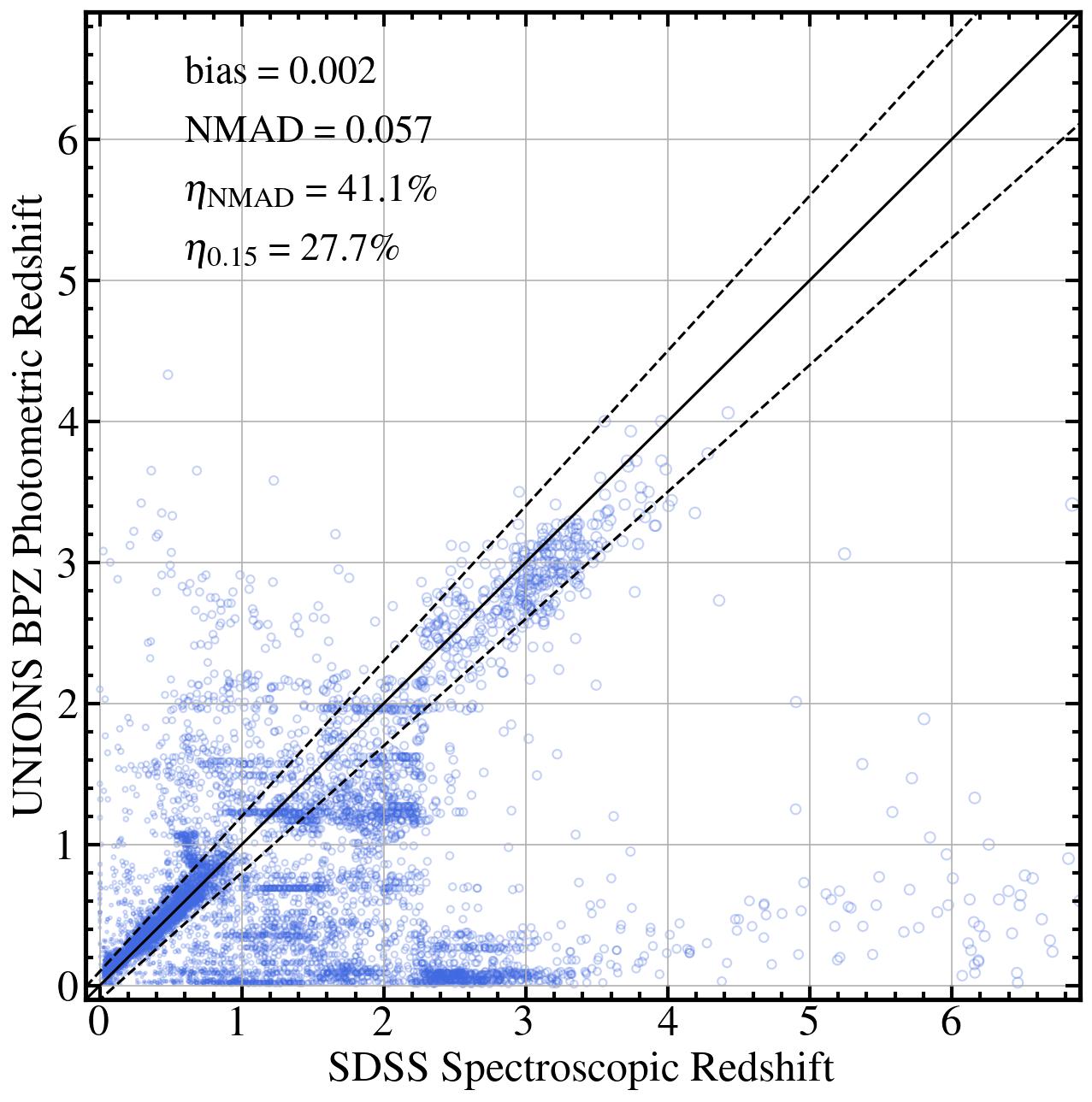}
\caption{Comparisons between UNIONS photometric and SDSS spectroscopic redshifts for $\upda{12,798}$ UNIONS-VLASS sources.
The solid line indicates $\tnr{z}{photo}=\tnr{z}{spec}$ while the dashed lines indicate $(\tnr{z}{photo}-\tnr{z}{spec})/(1+\tnr{z}{spec})=\pm0.1$, which represents the photo-$z$ scatter for guidance.
\label{fig: photo-z vs spec-z}}
\end{figure}

In Figure~\ref{fig: photo-z vs spec-z}, we compare the valid $\tnr{z}{photo}$ of our UNIONS-VLASS sources with the $\tnr{z}{spec}$ collected from SDSS DR16 \citep{ahumada_16th_2020} to check the reliability of the photo-$z$estimates, and $\upda{12,798}$ sources are included in this comparison.
The bias of the photo-$z$ estimates of these sources is $\upda{\sim0.002}$, and this bias remains for those at $\tnr{z}{photo}<1$.
We calculate the MAD of $\frac{\Delta z}{1+z}$ and define outliers according to the NMAD.
For all with $\tnr{z}{spec}$, the MAD is $0.038$, corresponding to bias $\upda{\sim0.002}$, and correspondingly, the NMAD is $\upda{1.4826 \times 0.038\approx0.057}$.
For those at $\tnr{z}{photo}>1$, the bias worsens to $\upda{0.026}$ and the scatter worsens to $\upda{0.213}$.
Adopting the criterion of $\frac{\Delta z}{1+z}>\mathrm{NMAD}\ (=0.057)$, $\upda{5264\ (41.1\%)}$ are outliers.
If choosing $\frac{\Delta z}{1+z}>0.15$ instead, 3540 (27.7\%) are outliers, which is still higher than that of UNIONS \citep{gwyn_unions_2025}.
This large fraction of outliers is likely attributed to sources whose spectroscopic redshifts are much larger than photometric ones, namely, $|\tnr{z}{spec}-\tnr{z}{photo}|>0.5$, occupying 23.3\% ($\upda{2981}$) of the total sample.
These large discrepancies are much more severe at $\tnr{z}{photo}<1.5$.
As we will discuss in \S\,\ref{sec: dropouts}, in which we select high-$z$ RLAGNs via the dropout-selection method, this could suggest an underestimate of the number density of high-$z$ sources in UNIONS.
\apjsa{This systematic underestimation in redshift is probably due to the absence of AGN-dominated templates in the photo-\textit{z} SED fitting procedure used by UNIONS \citep{gwyn_unions_2025}.
As a consequence, these RGs with large discrepancies between $\tnr{z}{photo}$ and $\tnr{z}{spec}$ are poor fits by the galaxy templates, because radio-selected sources can exhibit diverse optical properties -- ranging from low-luminosity radio AGNs in the nearby Universe to high-$z$ luminous RL quasars that outshine their host galaxies \citep{duncan_all-purpose_2022}.}
Nonetheless, we see good agreement between $\tnr{z}{spec}$ and $\tnr{z}{photo}$ at $z\gtrsim2.5$.
Therefore, our UNIONS-VLASS catalog provides a promising exploration for high-\textit{z} RLAGNs.

\begin{figure}[tp!]
\includegraphics[width=\columnwidth]{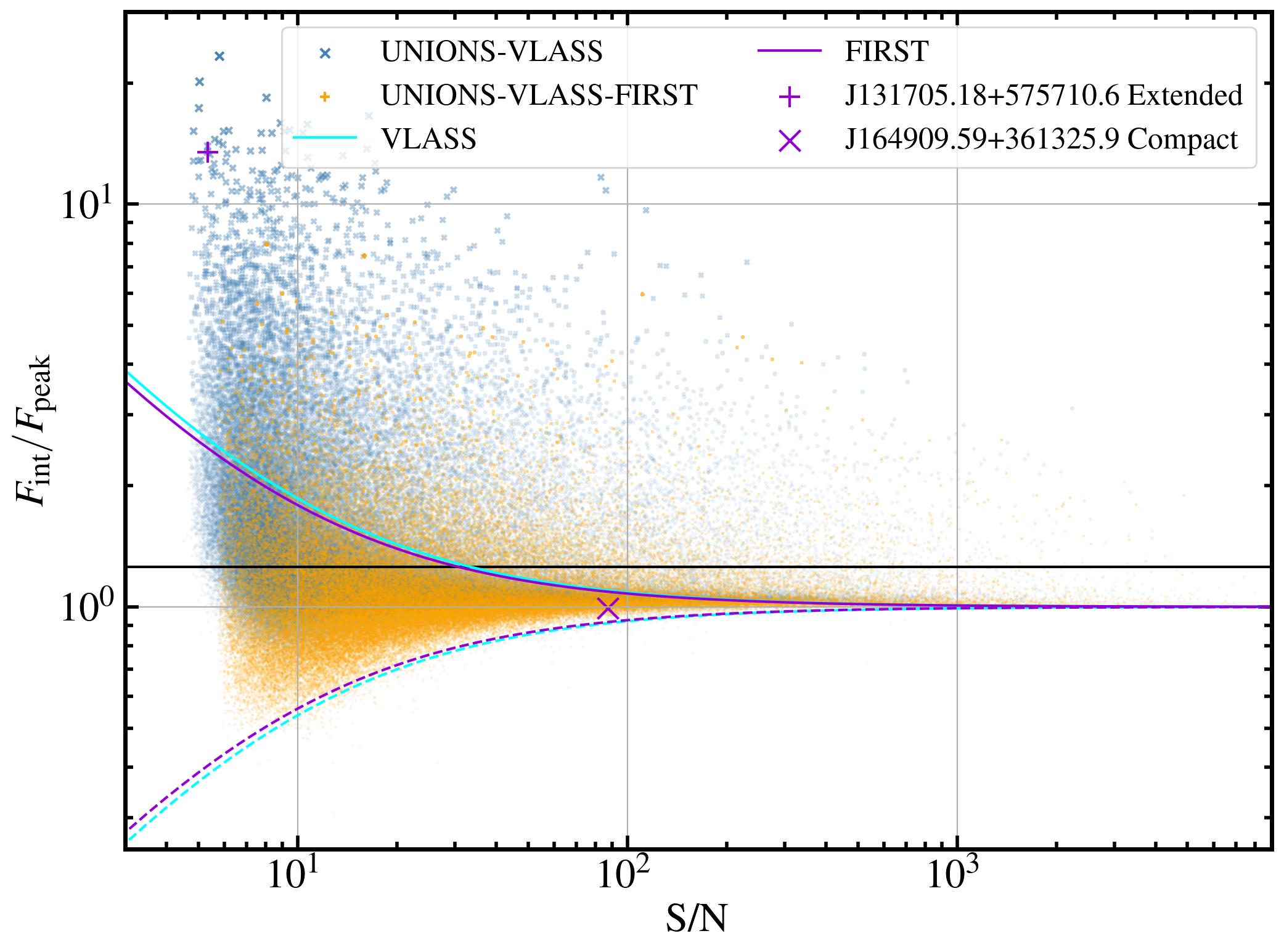}
\caption{The ratio of the total flux density to the peak flux density as a function of S/N defined as the ratio of the peak flux density to the survey detection limit for VLASS (blue crosses) and FIRST (orange pluses).
The dashed and solid lines represent the lower and upper envelopes, respectively (see text for details). 
Those lying above the upper envelope are considered extended sources.
The plus and crosses represent the examples of extended and compact sources, respectively, shown in Figure~\ref{fig: extended compact thumbnails}
\label{fig: extended compact sources}}
\end{figure}

\begin{figure}[tp!]
\includegraphics[width=\columnwidth]{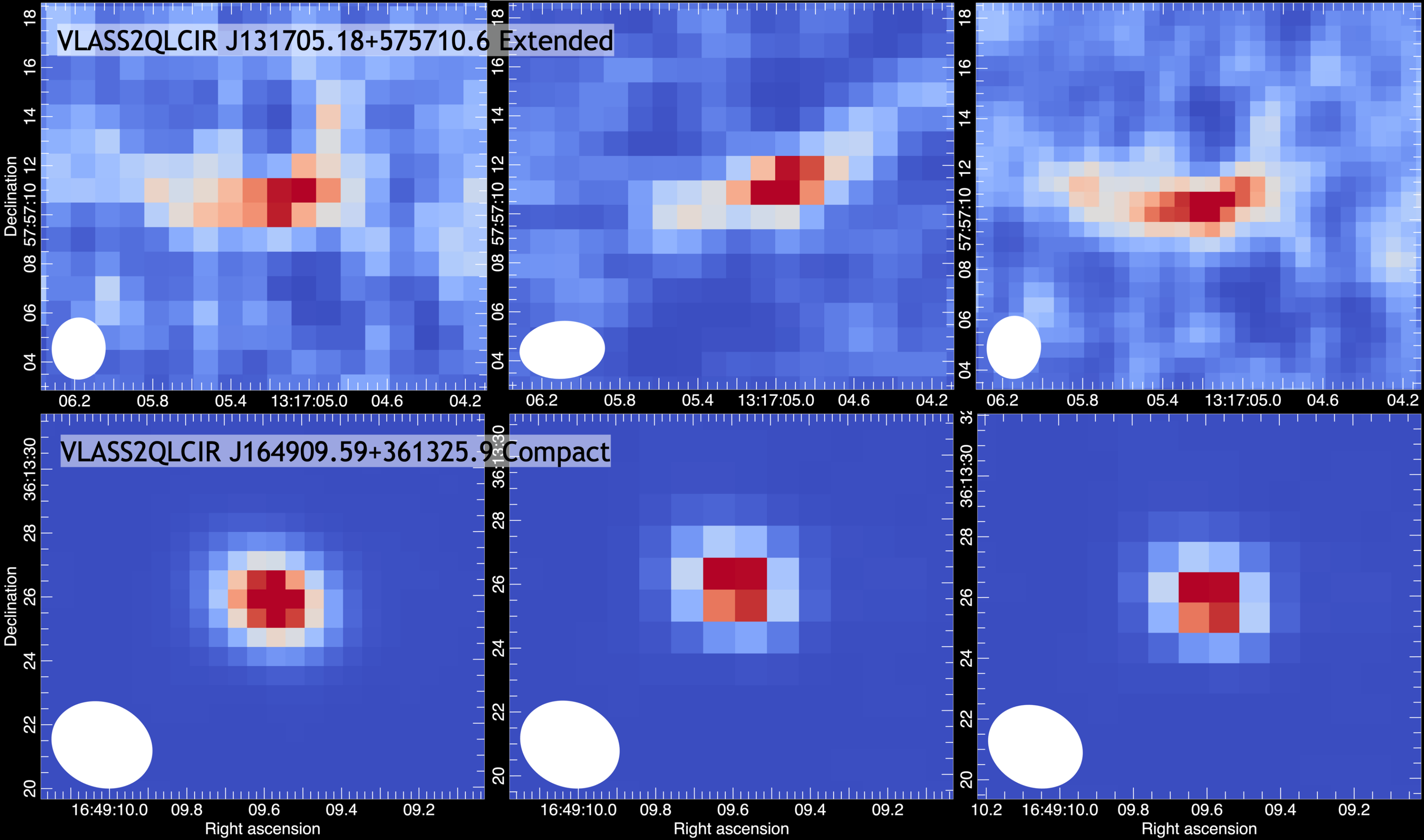}
\caption{We show examples of extended (upper)
and compact sources (lower) selected from the criterion described in \S\,\ref{subsec:extended sources} to demonstrate the validity.
The corresponding positions of these two sources $\tnr{F}{int}/\tnr{F}{peak}$ and S/N are shown in Figure~\ref{fig: extended compact sources}.
The white ellipses in the bottom left corner of each panel indicate the beam size.
For both rows, from left to right, the cutout is the product of VLASS Epoch 2 QL, Epoch 3 QL, and SE image, respectively.
\label{fig: extended compact thumbnails}}
\end{figure}

\subsubsection{Extended Radio Sources} \label{subsec:extended sources}
The morphology of radio emission, such as its compactness, presumably originating from radio jets, provides key information about the evolution of RLAGNs.
To distinguish between compact (unresolved) and extended (resolved) sources, we make use of the ratio of the total radio flux density ($\tnr{F}{int}$) to the peak radio flux density ($\tnr{F}{peak}$), a widely adopted method \citep[e.g.][]{ivezic_optical_2002,bondi_vla-virmos_2003,schinnerer_vla-cosmos_2007}.
This ratio directly measures the extension of a radio source, and for unresolved or marginally resolved sources, $\tnr{F}{int}=\tnr{F}{peak}$ \citep[e.g.,][]{bondi_vla-cosmos_2008}.
Extended and resolved sources have $\tnr{F}{int}>\tnr{F}{peak}$ whereas the background noise could lower the total flux density.
Therefore, we plot $\tnr{F}{int}/\tnr{F}{peak}$ as a function of the signal-to-noise ratio $\mathrm{S/N}\ (=\tnr{F}{peak}/\mathrm{rms})$ in Figure~\ref{fig: extended compact sources} for 146,212 ``clean'' UNIONS-VLASS sources (blue crosses) and those (79,638) with FIRST detections (orange dots).
\apjsa{Values of $\tnr{F}{int}/\tnr{F}{peak}<1$ are then attributed to the measurement uncertainties probably arising from the background noises that could lower $\tnr{F}{int}$ \citep[e.g.,][]{bondi_vla-virmos_2003}.}

To select extended sources, we first determine a lower envelope that encloses 99\% of the UNIONS-VLASS sources with $\tnr{F}{int}<\tnr{F}{peak}$, as indicated by the dashed cyan curve.
We then mirror it above the horizontal line defined by $\tnr{F}{int}/\tnr{F}{peak}=1$, and this upper envelope (solid cyan curve) is represented as:
\begin{equation}
    \tnr{F}{int}/\tnr{F}{peak}=1+8.6\times\lrp{\tnr{F}{peak}/\mathrm{rms}}^{-1}.
\end{equation}
Adopting these criteria, 29,862 ($\sim20\%$) of our UNIONS-VLASS sources are extended.
\apjsa{We perform a similar evaluation for sources detected in FIRST, where the FIRST S/N is defined as $(\tnr{F}{peak}-0.25)/\rm{rms}$ to account for the CLEAN bias \citep{helfand_last_2015}. 
The corresponding upper envelope is given by $\tnr{F}{int}/\tnr{F}{peak}=1+7.9\times\left[\lrp{\tnr{F}{peak}-0.25}/\mathrm{rms}\right]^{-1}$, which is generally in line with $\tnr{F}{int}/\tnr{F}{peak}=1+6.5\times\lrp{\tnr{F}{peak}/\mathrm{rms}}^{-1}$ derived for the HSC-FIRST catalog \citep[][]{yamashita_wide_2018}.}

Alternatively, thanks to the higher sensitivity and better resolution, the VLA-COSMOS 3~GHz survey \citep{smolcic_vla-cosmos_2017} determined the envelope by a 95\% enclosure of the sources below the $\tnr{F}{int}=\tnr{F}{peak}$ line.
Adopting this criterion, the upper envelope is 
\begin{equation}
    \tnr{F}{int}/\tnr{F}{peak}=1+3.7\times\lrp{\tnr{F}{peak}/\mathrm{rms}}^{-1},
\end{equation}
and 63,203 ($\sim43\%$) of our UNIONS-VLASS sources are extended.
In this work, we adopt Eq.~(1) as the threshold considering the similar beam sizes and sensitivities between VLASS and FIRST, and extended radio sources are labeled with $\mathtt{1}$ in column $\mathtt{extended}$.
Examples of extended and compact radio sources are shown in Figure \ref{fig: extended compact thumbnails} and they are marked on the $\tnr{F}{int}/\tnr{F}{peak}-\mathrm{S/N}$ plane (Figure \ref{fig: extended compact sources}).

\subsubsection{Additional Columns}
In the final UNIONS-VLASS radio galaxy catalog, we provide $\mathtt{Z\_B\_ref}$ for the preferred UNIONS photo-$z$.
If $\mathtt{flag\_z\_photo==0}$, $\tnr{z}{photo}$ of $\mathrm{UNIONS5000}_{ugriz}$ is used; if $\mathtt{flag\_z\_photo==1}$, $\tnr{z}{photo}$ of $\mathrm{UNIONS5000}_{ugri}$ is used; and if $\mathtt{flag\_z\_photo==-1}$, the photometric redshifts are those from $\mathrm{UNIONS5000}_{ugriz}$ but should not be trusted.
We also provide $\mathtt{z\_ref}$ for the preferred redshift, and the SDSS spectroscopic redshifts are used if available; otherwise $\mathtt{Z\_B\_ref}$ is used.

We calculate the angular separations between each pair of UNIONS, VLASS, FIRST, and LoTSS sources, and they are recorded in the corresponding columns starting with $\mathtt{sep\_\{survey1\}\_\{survey2\}}$.
We calculate the spectral indices between each pair of radio surveys if there are detections, and the corresponding columns start with $\mathtt{alpha\_\{frequency1\}\_\{frequency2\}}$.
\apjsa{Since the spectral index is sensitive to uncertainties in the flux densities, we provide the corresponding uncertainties in the spectral index in columns $\mathtt{alpha\_\{frequency1\}\_\{frequency2\}\_err}$.}

We calculate the spectral luminosities [$\whz$] at 1.4 and 3~GHz using the following equation:
\begin{equation}
    L_\nu=\frac{S_\mathrm{obs}4\pi D^2_L}{(1+z)^{1+\alpha}},
\end{equation}
where $S_\mathrm{obs}$ [mJy] is the integrated flux density, $D_L$ is the luminosity distance, $z$ is the redshift, and $\alpha$ is the spectral index defined in the positive convention $S_\mathrm{obs} \propto \nu^{\alpha}$.
The redshift adopts those stored in $\mathtt{`Z\_B\_ref'}$ representing the UNIONS photometric redshifts.
For $\tnr{L}{3GHz}$, if the UNIONS-VLASS source is detected in FIRST, then $\mathtt{alpha\_1400\_3000}$ is used, otherwise $\tnrd{\alpha}{3GHz}{1.4GHz}$ is assumed to be $-0.7$ \citep{condon_radio_1992,dermer_high_2009}.
For $\tnr{L}{1.4GHz}$, we adopt the following strategy for a UNIONS-VLASS source if it is:
\begin{itemize}
\item  detected in FIRST but not in LoTSS: $S_\mathrm{obs}=S_\mathrm{FIRST}$ and $\tnrd{\alpha}{1400 MHz}{144MHz}=-0.7$ used \citep{gasperin_radio_2018}; \\
\item detected in FIRST and LoTSS: $S_\mathrm{obs}=S_\mathrm{FIRST}$ and $\mathtt{alpha\_144\_1400}$ used;
\item not detected in FIRST and LoTSS: the total flux density of VLASS is scaled to that at 1.4~GHz assuming $\tnrd{\alpha}{3GHz}{1.4GHz}=-0.7$, and then $\tnr{L}{1.4GHz}$ is calculated further assuming $\tnrd{\alpha}{1400 MHz}{144MHz}=-0.7$.
\end{itemize}

The spectral luminosities will be discussed in detail in \S\,\ref{sec: radio loudness}.
The calculations and discussions of radio AB magnitudes ($\mathtt{mag\_\{survey\}}$) and radio loudness ($\mathtt{r\_obs}$) will be presented in \S\,\ref{sec: radio source counts} and \S\,\ref{sec: radio loudness}, respectively. 
A full list and details of the columns are given in Table~\ref{tab: catalog columns}.

\begin{figure*}[htp!]
\includegraphics[width=\textwidth]{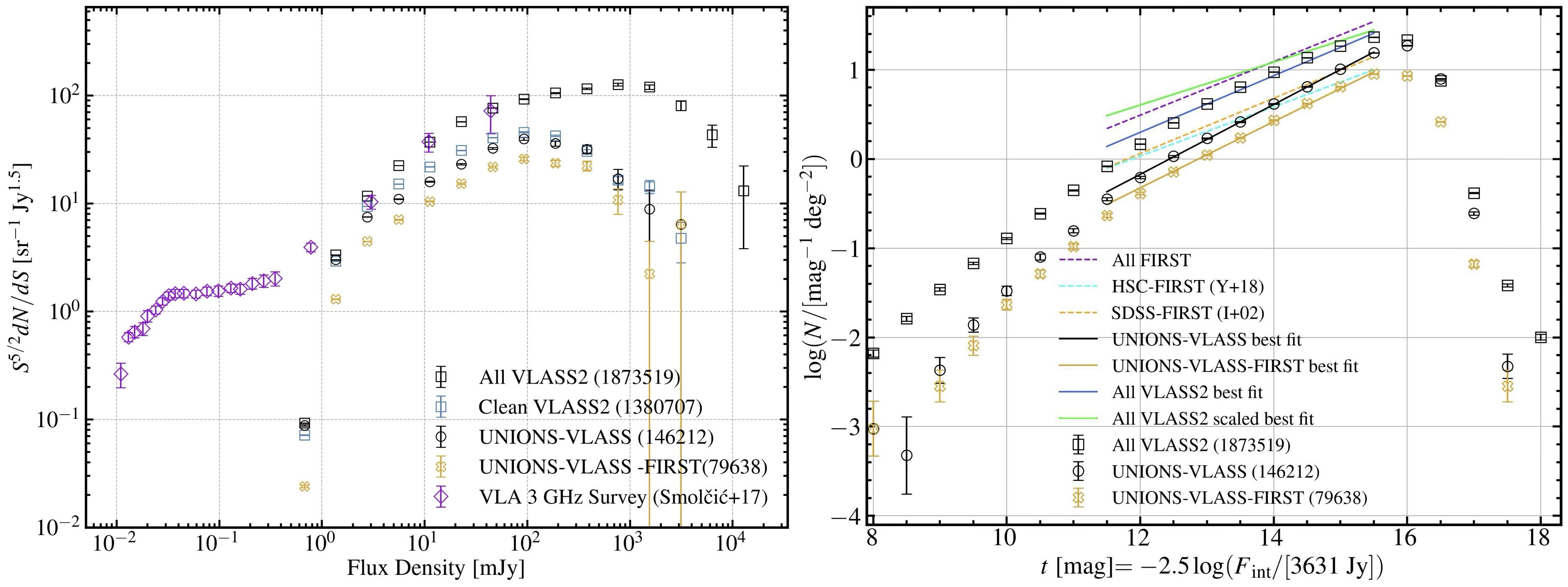}
\caption{
\apjsb{\textit{Left}: Radio number counts normalized to the static Euclidean geometry.}
\textit{Right}: VLASS 3~GHz source counts as a function of the AB radio magnitude \textit{t}. 
Source counts at $11.5<t<15.5$ are fitted with a linear function, and the best fits are indicated by the colored line, while the solid ones are fittings in this work and dashed ones are from the literature.
The black circles represent UNIONS-VLASS sources, black squares represent the ``all'' VLASS Epoch 2 catalog with criteria (except $\mathtt{NN\_dist}>30\arcsec$) in \S\,\ref{sec: vlass} applied, and golden crosses represent UNIONS-VLASS sources detected in FIRST, and the size of each sample is shown in the parenthesis. 
Poisson errors in the sampling bins are represented by vertical bars.
\label{fig: radio number counts}}
\end{figure*}

\section{Radio and Optical Number Counts} \label{sec:properties}
We here summarize the number counts and basic radio properties of UNIONS-VLASS radio galaxies, with comparisons from previous radio galaxy catalogs in the literature.

\subsection{UNIONS-VLASS 3~GHz Radio Source Counts} \label{sec: radio source counts}
Radio source number counts are often normalized to Euclidean geometry and examined on the plane of flux density ($S$) and $S^{2.5}\ dN/dS$, where $N$ is the number of sources within each flux density bin (left panel of Figure~\ref{fig: radio number counts}).
\apjsb{Compared to the VLA-COSMOS 3~GHz Survey that reaches a median sensitivity of 2.3 $\mathrm{\mu Jy\ beam^{-1}}$ (\citealt{smolcic_vla-cosmos_2017}; indicated by violet diamonds), VLASS is limited in probing radio sources with $\tnr{F}{int}<1$~mJy.
As a result, VLASS is comparatively biased toward brighter sources and is therefore more likely to select RL populations, which will be further discussed in \S\,\ref{sec: radio loudness}.}

We use also the plane of the radio AB magnitude \textit{t} and the number counts \textit{N} in each magnitude bin to make direct comparisons with FIRST catalogs at 1.4~GHz.
Here, the radio AB magnitude is defined as:
\begin{equation}
    t\ [\mathrm{AB\ mag}]=-2.5\log\lrp{\tnr{F}{int}/3631\ \mathrm{Jy}}.
\end{equation}
The $3\sigma\ (=0.36$ mJy) detection limit of VLASS then corresponds to 17.5 and the $5\sigma\ (=1$ mJy) detection limit of FIRST then corresponds to 16.4.
In the right panel of Figure~\ref{fig: radio number counts}, we show radio source counts for 146,212 ``clean'' UNIONS-VLASS radio galaxies (black circles).
We make a UNIONS-VLASS subsample for 79,638 sources that are also detected in FIRST (golden crosses) and compare it with 1.4~GHz radio galaxy catalogs employing FIRST \citep{ivezic_optical_2002,yamashita_wide_2018}.
We also apply the criteria described in \S\,\ref{sec: vlass}, except $\mathtt{NN\_dist>30\arcsec}$, to the entire VLASS2 catalog, resulting in 1,873,519 sources (black squares).
We compare this ``all'' VLASS2 sample with the ``clean'' VLASS2 sample that contains 1,380,707 sources used for the cross-match.
The distributions of these samples at $11.5<t<15.5$ are fitted with a linear function and are indicated by the solid lines, and the best-fit parameters are summarized in Table \ref{tab: fits}.
The best fits of the data collected from the literature are indicated by the dashed lines.

The trends between ``all'' VLASS (blue solid line) and UNIONS-VLASS sources (black solid line) are in rough consistency.
The global difference and the larger discrepancy at the bright end are attributed to the fact that we have applied $\mathtt{NN\_dist>30\arcsec}$ to VLASS2 in the pre-cross-match stage.
This criterion may remove a significant fraction of low-$z$ sources that have bright radio lobes and hotspots identified as adjacent components, contributing to the number counts at the bright end.
If we compare UNIONS-VLASS with the ``clean'' VLASS2 used for the cross-match, then this discrepancy diminishes, and the global difference is consistent with the matching rate of $0.594$ (left panel of Figure~\ref{fig: radio number counts}).

Compared with 1.4~GHz HSC- and SDSS-FIRST source counts (golden and cyan dashed lines), our 3~GHz UNIONS-VLASS sources (black circles) have a slightly steeper increase at $11.5<t<15.5$.
This could be explained by our choice of $\mathtt{NN\_dist>30\arcsec}$.
The best fit of the subsample of our UNIONS-VLASS sources with FIRST detections (golden solid line) has a slope in general consistency with previous studies.
This comes from the fact that the sky coverage of UNIONS-VLASS is beyond the survey area of FIRST, thus resulting in an underestimate of the real number counts for our UNIONS-VLASS-FIRST sample.

We further compare the number counts between VLASS2 and FIRST (dashed violet line).
Since VLASS2 and FIRST are operated at different frequencies, we scale the 3~GHz flux densities of the ``all'' VLASS2 adopting $\alpha\mathrm{^{FIRST}_{VLASS}}=-0.7$, and the best fit of VLASS2 scaled is indicated by the green solid line.
The global amplitude of the scaled ``all'' VLASS is elevated, whereas the slope becomes flatter.
On the other hand, although the ``clean'' VLASS2 is in general sublinear to FIRST, these two surveys share a consistent trend.
These results clearly reflect the complexity of spectral indices of real radio sources that they cannot always be described by a simple power-law distribution (see \S\,\ref{subsubsec:radio color-color} for details).

\begin{figure}[tp!]
\includegraphics[width=\columnwidth]{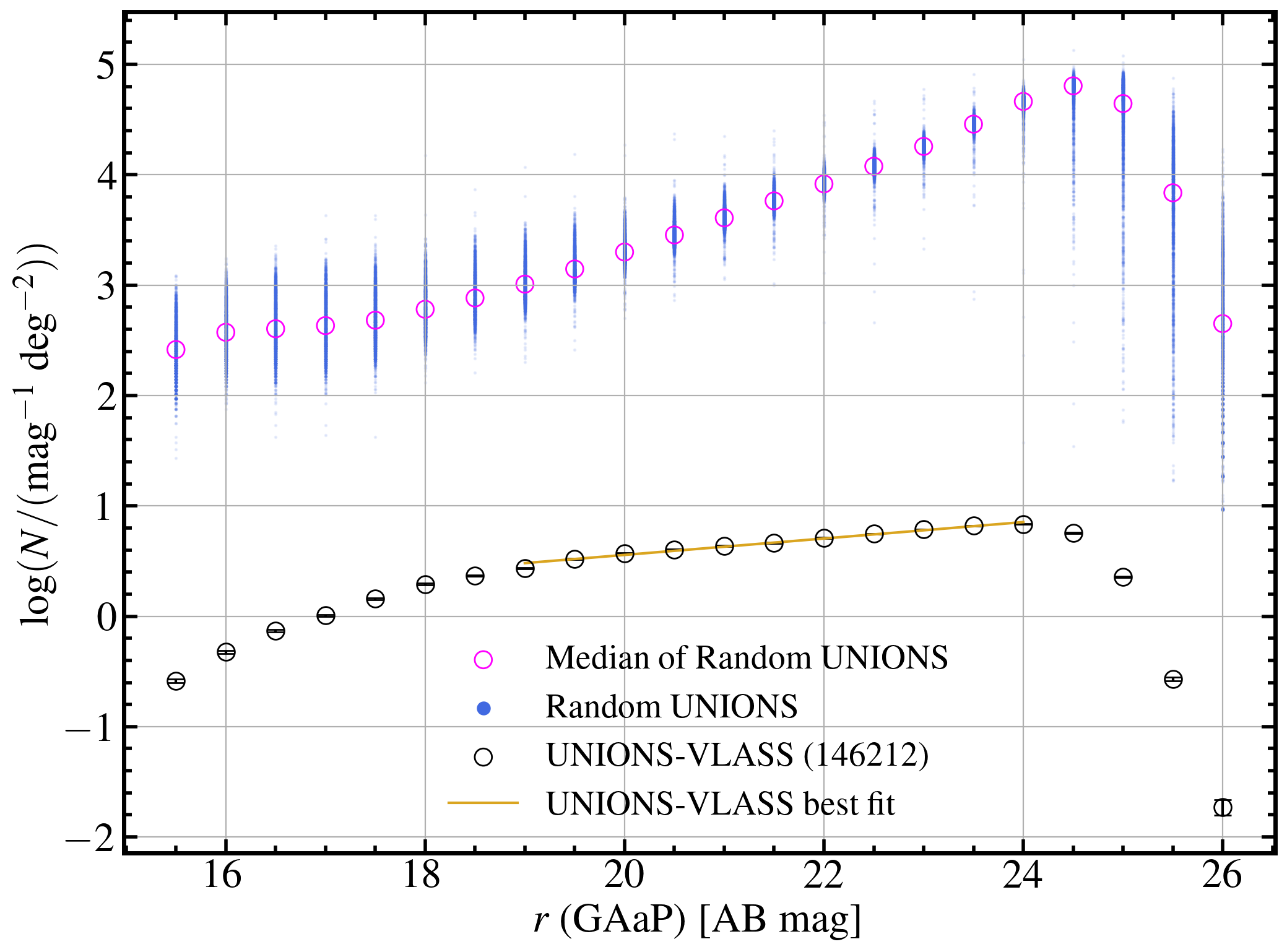}
\caption{Optical UNIONS $r$-band number counts of 146,212 UNIONS-VLASS radio galaxies with valid $r$-band photometry.
The blue dots show distributions of sources in each of the randomly selected 2,000 UNIONS tiles, the magenta circles indicate the median of 2,000 UNIONS tiles at each photometry bin, and the black circles represent UNIONS-VLASS sources.
The Poisson errors in the sample are represented by vertical bars.
The gold solid line indicates the best fit for UNIONS-VLASS sources.
\label{fig: optical number counts r}}
\end{figure}

\begin{table}[tp!]
\caption{Fitting Results of Radio and Optical Number Counts}
\begin{center}
\label{tab: fits}
\begin{tabular}{lcc}
\hline
 & $a$ & $b$ \\
\hline
All VLASS & $-3.509\ (\pm0.269)$ & $0.317\ (\pm0.018)$ \\
UNIONS-VLASS & $-4.870\ (\pm0.067)$ & $0.391\ (\pm0.005)$ \\
UNIONS-VLASS-FIRST & $-4.779\ (\pm0.142)$ & $0.371\ (\pm0.010)$ \\
\hline
$r$-band & $-0.932\ (\pm0.067)$ & $0.074\ (\pm0.003)$ \\
\hline
\end{tabular}
\end{center}
{
The best fits of radio source counts in Figure~\ref{fig: radio number counts} and optical \textit{r-}band source counts in Figure~\ref{fig: optical number counts r}.
The fitting function is $\log(N/[\mathrm{mag^{-1}\ deg^{-2}}])=a+bt\ \mathrm{[AB\ mag]}$. 
The numbers in the parentheses represent $1\sigma$ uncertainties in the fitting results.
}
\end{table}

\subsection{UNIONS-VLASS Optical Source Counts} \label{subsec: optical source counts}
We investigate the differential counts in $r$-band for our 146,212 ``clean'' UNIONS-VLASS sources, as shown in Figure~\ref{fig: optical number counts r} by black circles.
We also plot the optical counts of UNIONS by randomly selecting 2000 tiles in the NGP.
The fall-offs at the faint end ($\tnr{r}{ABmag}\geq24$) of UNIONS-VLASS and UNIONS are highly consistent, suggesting that the limitation in associating radio galaxies with faint host galaxies is the depth of the optical survey.
\apjsb{At the bright end ($\tnr{r}{ABmag}\lesssim17$), the optical number counts of UNIONS-VLASS sources decrease with the brighter $r$-band magnitude. 
This fall-off, representing the saturation effect and intrinsic rarity of quasars \citep{yamashita_wide_2018}, is not observed in the UNIONS galaxies.
Therefore, this declining trend in UNIONS-VLASS is likely to be the result of the elimination of the brightest nearby radio sources in ``clean'' VLASS2.}
For UNIONS-VLASS sources with $19\leq \tnr{r}{ABmag}\leq24$, the best fit is indicated by the golden solid line.
The source count has a weak dependency on the $r$-band magnitude throughout this range, despite an increasing completeness of UNIONS along with the decreasing $r$-band brightness.
This implies a decreasing detectability of radio galaxies in the UNIONS catalog with an increasing apparent optical magnitude: from $0.26\%$ at $r=19$ to $0.01\%$ at $r=24$.
\apjsa{This trend is also confirmed by VLASS RGs in Euclid Deep Field North and Fornax, where $\tnr{f}{matches}$ reaches $\sim60\%$ down to a VIS magnitude of 24 (Zhong et al. in prep.).
Such a decline in detectability may result from a combination of factors: (1) the Malmquist bias \citep{miley_distant_2008}, as the detection limit of VLASS increasingly excludes faint sources at higher redshifts; and (2) the intrinsic rarity of radio AGNs at high-$z$, partly due to the rising energy density of the cosmic microwave background, which enhances inverse Compton scattering of synchrotron-radiating electrons, thereby exacerbating the selection bias \citep{condon_essential_2016,saxena_modelling_2017}.}

\begin{figure*}[tp!]
\includegraphics[width=\textwidth]{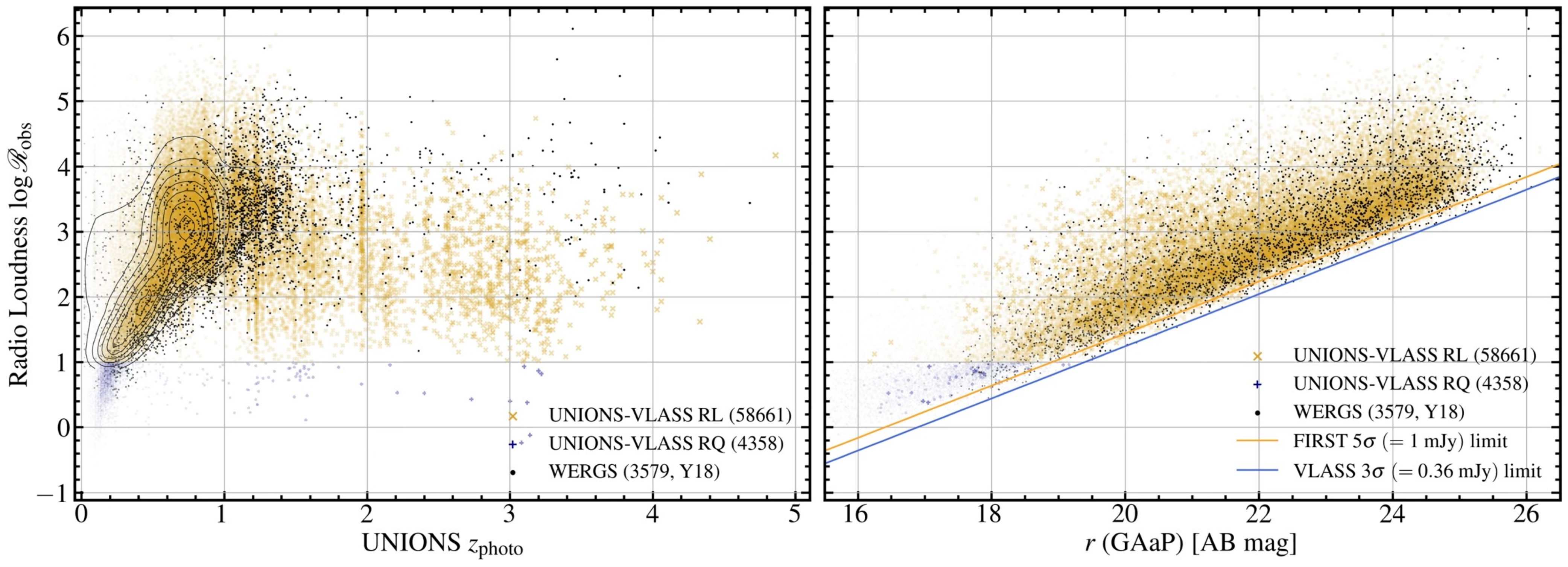}
\caption{\textit{Left}: observed radio loudness as a function of the redshift.
\textit{Right}: observed radio loudness as a function of the $r$-band magnitudes.
These distributions include \upda{63,019} UNIONS-VLASS radio galaxies with valid $r$-band photometric measurements.
The observed radio loudness $\robs$ is defined as the ratio of the flux density at 1.4~GHz to that of the $r$-band.
The RLAGNs are characterized by $\log\robs>1$ and are indicated by orange crosses, while the dark violet pluses indicate RQ AGNs.
The contours in the left panel are only for RL populations.
The black dots represent 3,579 radio galaxies from the HSC-FIRST catalog \citep{yamashita_wide_2018}.
We also show the detection limits of FIRST and VLASS by orange and blue solid lines, respectively.
\label{fig: radio loudness}}
\end{figure*}

\begin{figure*}[tp!]
\includegraphics[width=\textwidth]{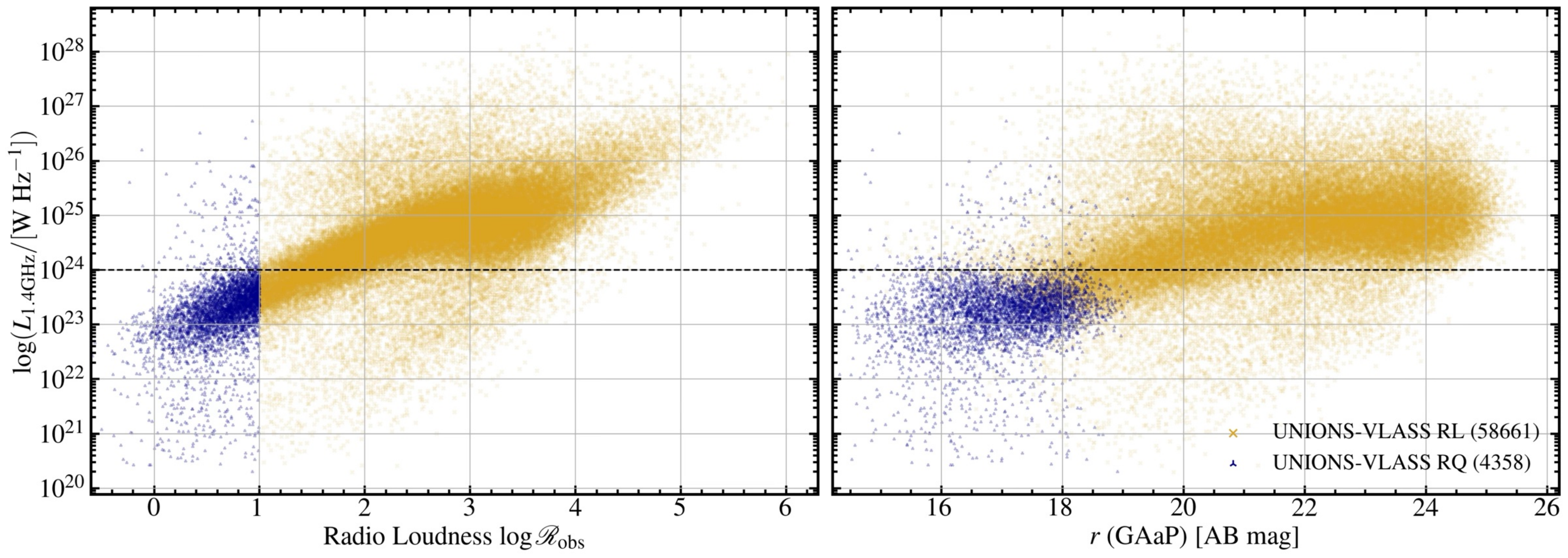}
\caption{
\textit{Left}: 1.4~GHz spectral luminosity as a function of the observed radio loudness.
\textit{Right}: 1.4~GHz spectral luminosity as a function of the $r$-band magnitudes.
These distributions include \upda{63,019} UNIONS-VLASS radio galaxies with valid $r$-band photometric measurements.
The horizontal black dashed line indicates the RL/RQAGN demarcation of $\tnr{L}{1.4GHz}\sim10^{24}\whz$ \citep{ivezic_optical_2002}.
The colored symbols are the same as Figure~\ref{fig: radio loudness}.
\label{fig: radio loudness vs luminosity}}
\end{figure*}

\begin{figure*}[htp!]
\includegraphics[width=\textwidth]{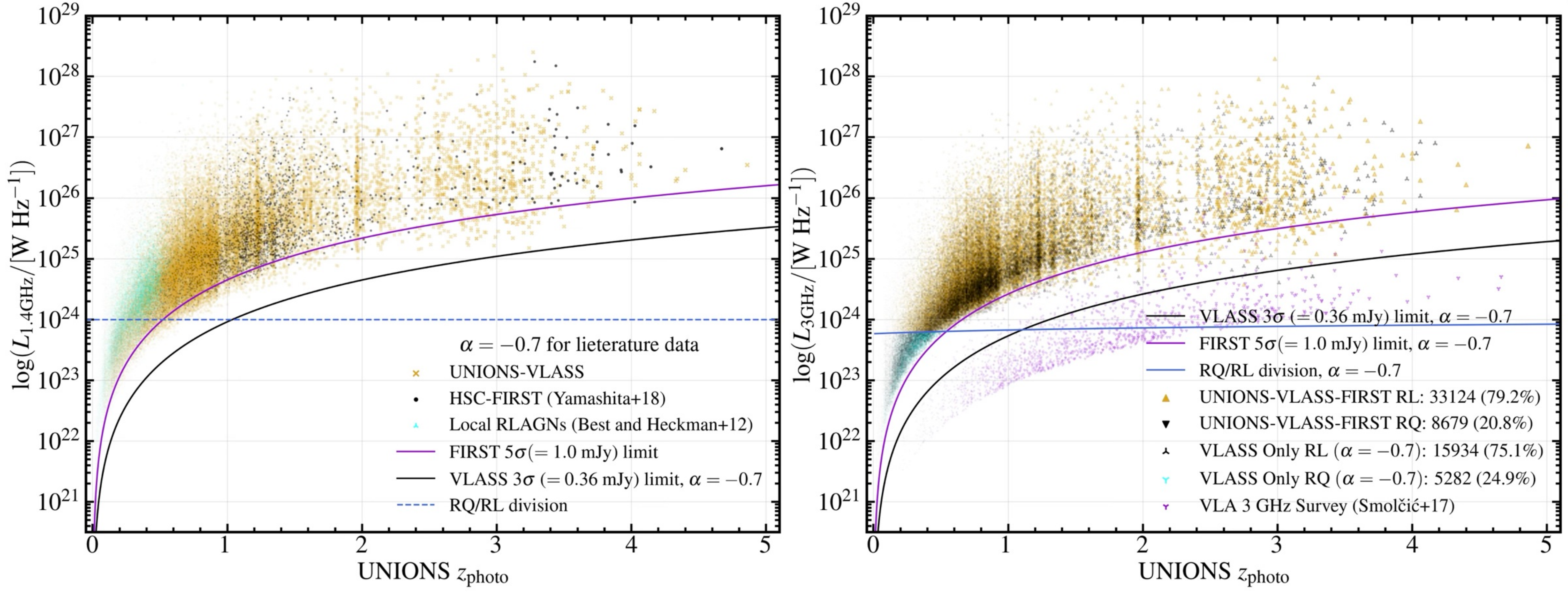}
\caption{
\textit{Left}: The 1.4~GHz spectral luminosity as a function of photo-\textit{z} for $\upda{63,019}$ UNIONS-VLASS radio galaxies (orange crosses), in comparison with HSC-FIRST AGNs (black dots) and local RLAGNs (cyan triangles).
VLASS flux densities at 3~GHz are converted to those at 1.4~GHz, adopting $\tnrd{\alpha}{VLASS}{FIRST}$ for those detected in FIRST and adopting $\alpha=-0.7$ for those without FIRST detection.
\textit{Right}: The 3~GHz spectral luminosity as a function of photo-\textit{z} for UNIONS-VLASS sample, in comparison with VLA 3~GHz survey.
We separate $\upda{63,019}$ UNIONS-VLASS radio galaxies with valid $\tnr{z}{photo}$ into two subsamples.
$\tnr{L}{3GHz}$ of those detected in both VLASS and FIRST ($\upda{41,803}$ in total) are calculated based on $\alpha\mathrm{_{VLASS}^{FIRST}}$, and of those without FIRST detection ($\upda{21,216}$ in total) are calculated using $\alpha=-0.7$.
These sources are further divided into RL and RQ populations following the division line $\tnr{L}{1.4GHz}\sim10^{24}\whz$ \citep{ivezic_optical_2002}, which is indicated by the blue solid line.
In both panels, the violet line indicates the $5\sigma\ (=1.0$ mJy) detection limit of FIRST and the black one indicates the $3\sigma\ (=0.36$ mJy) detection limit of VLASS.
\label{fig: radio luminosity}}
\end{figure*}

\subsection{Radio-Loud AGNs} \label{sec: radio loudness}
\subsubsection{Radio Loudness}
To separate RLAGNs from radio-quiet (RQ) populations, the observed radio loudness $\robs$ is often used.
A widely adopted quantification of $\robs$ is defined as the radio-to-optical flux ratio where no $k-$corrections are applied:
\begin{equation}
    \log\robs=\log\lrp{\tnr{F}{radio}/\tnr{F}{optical}}=0.4(m-t),
\end{equation}
where $m$ is one of the optical magnitudes and $t$ is the AB radio magnitudes in Eq.~(3).
Although the \textit{i-}band photometry is commonly chosen for $\tnr{F}{optical}$, the fact that UNIONS5000 is built upon the $r$-band detections ensures a hundred percent completeness of $\robs$ for our catalog.
We thus use the $r$-band photometry instead.
The median difference between $r$- and $i$-band photometry, as shown in the left panel of Figure~\ref{fig: unions2000 vs 5000}, is 0.44 in both UNIONS5000 and UNIONS2000 data releases.
Therefore, statistically, the choice of $\tnr{r}{ABmag}$ has insignificant impacts on the final classifications.
For $\tnr{F}{radio}$, we use the 1.4~GHz flux density.
The 1.4~GHz flux densities of UNIONS-VLASS sources with FIRST detection are calculated using FIRST flux densities.
Those not detected in FIRST have their VLASS 3~GHz flux densities scaled to those at 1.4~GHz using $\tnrd{\alpha}{3GHz}{1.4GHz}=-0.7$.
RLAGNs are then selected with $\log\robs>1$.
In this manner, for 146,212 sources in the ``clean'' UNIONS-VLASS catalog, 138,266 ($\sim94.6\%$) are RL and 7,946 ($\sim5.4\%$) are RQ populations.

\subsubsection{UNIONS-VLASS Subsample with Valid Photo-z}
To better understand RLAGNs across cosmic time, we focus on the subsample of $\upda{63,019}$ radio galaxies with valid UNIONS photo-$z$.
In this subsample, \upda{58,661} are RL and \upda{4,358} are RQ populations based on the $\robs$ criterion.
We show the distribution of $\robs$ as a function of the photometric redshift in the left panel of Figure~\ref{fig: radio loudness}.
Compared to the HSC-FIRST catalog that has $\log\robs\sim4$ at $\tnr{z}{photo}\sim2-4$ \citep{yamashita_wide_2018}, the bulk of our UNIONS-VLASS radio galaxies at this redshift range have $\log\robs\sim2-3$, and there is no evolution of $\robs$ within this redshift bin.
This may be explained by the fact that the HSC-FIRST catalog uses $\tnr{i}{ABmag}$ rather than $\tnr{r}{ABmag}$, and because of the greater depth of the HSC $i$-band, \citet{yamashita_wide_2018} captured more optically faint sources.
In the panel showing $\robs-\tnr{r}{ABmag}$ (right panel of Figure~\ref{fig: radio loudness}), we also plot the detection limit of VLASS and FIRST.
\apjsa{We see that at $\tnr{i}{ABmag}\gtrsim19.5$, all radio galaxies are radio-loud, and the RQ populations only exist at $\tnr{i}{ABmag}$ brighter than around 19, which clearly reflects the selection bias of RLAGNs based on $\robs$, that is, this criterion favors optically faint galaxies.}

To investigate whether the current abundance of RLAGNs in UNIONS-VLASS is due to the faintness of the optical counterparts, we adopt an alternative approach to selecting RLAGNs via the spectral luminosity threshold $\tnr{L}{radio}\sim10^{24}\whz$ established from FIRST RL/RQAGNs \citep{ivezic_optical_2002}.
This threshold depends on the reduced Hubble constant and we adopted $\tnr{h}{50}$ for consistency between works, and $\tnr{L}{radio}\sim2\times10^{24}\whz$ if $\tnr{h}{70}$ is used instead, which does not result in significant differences.
For $\upda{63,019}$ UNIONS-VLASS radio galaxies with valid $\tnr{z}{photo}$, $\upda{49,231}$ are RLAGNs with $\tnr{L}{radio}>10^{24}\whz$
There are $\upda{49,012}$ sources commonly identified as RLAGNs via selections based on $\tnr{L}{1.4GHz}$ and $\robs$.
Fractionally, the consistency is \upda{$\sim99.6\%$ (49,012/49,231)} for $\tnr{L}{1.4GHz}$ and \upda{$\sim83.6\%$ (49,012/58,661)} for $\robs$, respectively.

We show $\tnr{L}{1.4GHz}$ as a function of $\robs$ and $\tnr{r}{ABmag}$ in Figure~\ref{fig: radio loudness vs luminosity}.
It is found that RQAGNs, classified by $\log \robs < 1$, generally exhibit weaker $\tnr{L}{1.4,\mathrm{GHz}}$ values compared to RLAGNs.
Nonetheless, some RQAGNs are brighter than $\tnr{L}{1.4GHz}=10^{25}\whz$ with $\tnr{i}{ABmag}=16-18$, suggesting that they could be RL quasars.
There is a clear correlation between $\tnr{L}{1.4GHz}\ (\sim10^{23}-10^{24}\whz)$ and $\log\robs\ (\sim0-2)$.
Similarly, a weak correlation also exists for $\tnr{L}{1.4GHz}\ (\sim10^{23}-10^{24}\whz)$ and $\tnr{r}{ABmag}\ (\sim16-18)$.
These correlations are likely associated with the cosmic evolution of $\robs$ at $z<1$ (left panels of Figure~\ref{fig: radio loudness} and Figure~\ref{fig: imag, radio mag}).
On the other hand, beyond the RL-RQ demarcation of $\tnr{L}{radio}\sim10^{24}\whz$, the spectral luminosity has no evolution along $\tnr{r}{ABmag}\sim22-26$.

We further plot 1.4~GHz and 3~GHz spectral luminosities in the left and right panels of Figure~\ref{fig: radio luminosity}, respectively.
For those detected in FIRST, $\tnr{L}{3GHz}$ is calculated adopting $\tnrd{\alpha}{VLASS}{FIRST}$, while those without detection adopt a spectral index of $-0.7$.
We overlay the RL and RQ demarcation on the distribution of $\tnr{L}{1.4GHz}$ and scale it to that of $\tnr{L}{3GHz}$ adopting $\tnrd{\alpha}{3GHz}{1.4GHz}=-0.7$.
Limited by the sensitivity, both VLASS and FIRST cannot reach the depth required to detect RQAGNs at $z>1$.
Therefore, our UNIONS-VLASS radio galaxies $z>1$ are (almost) radio-loud, no matter whether they are identified based on $\log\robs>1$ or $\tnr{L}{1.4GHz}>10^{24}\whz$.

\begin{figure}[tp!]
\includegraphics[width=\columnwidth]{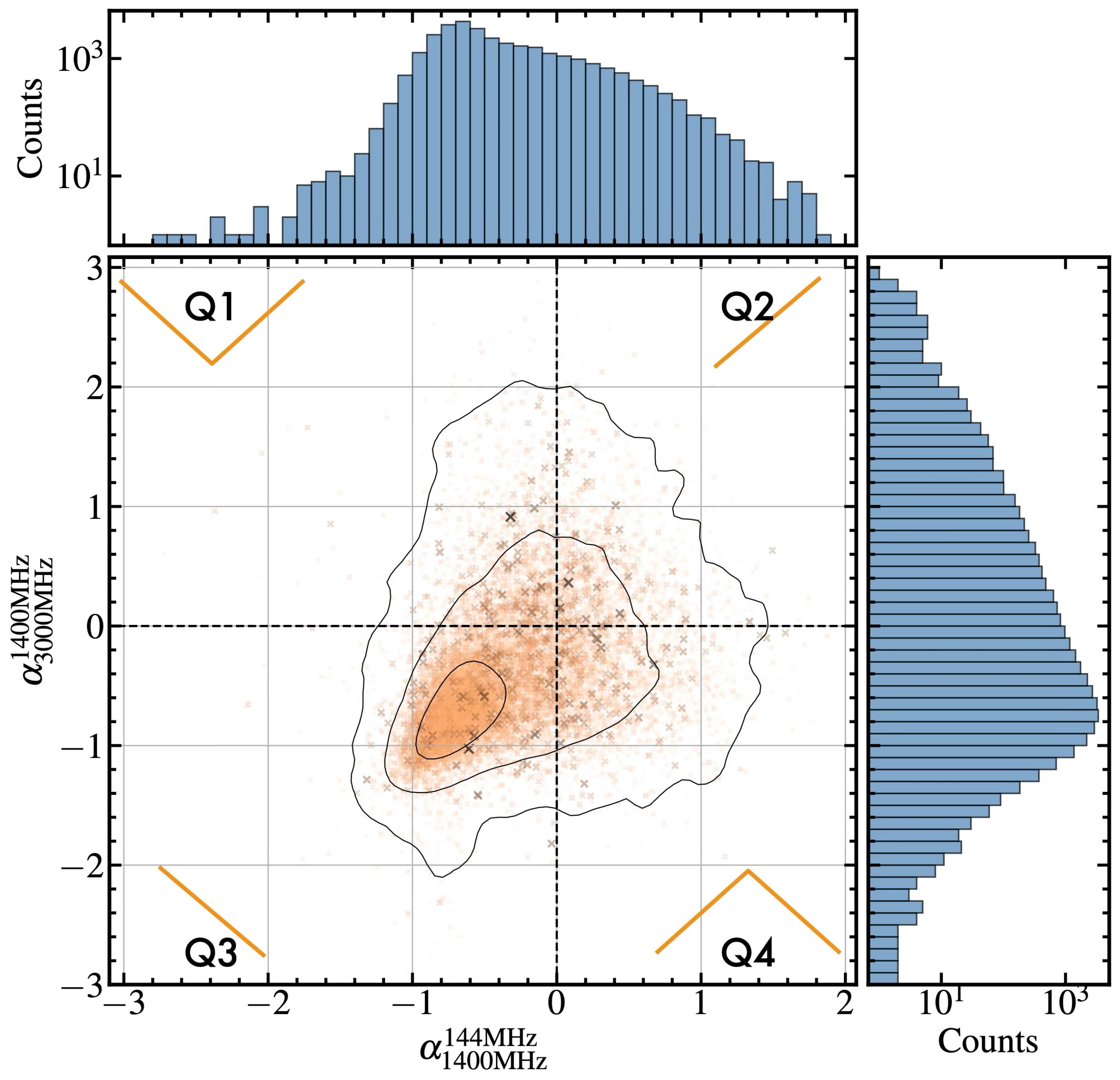}
\caption{The radio color-color diagram of $\tnrd{\alpha}{FIRST=1400MHz}{LoTSS=144Mhz}$ and $\tnrd{\alpha}{VLASS=3000MHz}{FIRST=1400Mhz}$ for \apjsa{30,091} UNIONS-VLASS sources with FIRST and LoTSS detections.
The darker and larger marker represents a higher redshift.
The spectral shapes are shown in the four quadrants, where Q1 represents upturned, Q2 represents inverted, Q3 represents power-law, Q4 represents peaked spectrum, respectively (see texts for details).
\label{fig: radio cc}}
\end{figure}

\section{Discussion} \label{sec:discussion}

\subsection{Radio Color-Color Diagram} \label{subsubsec:radio color-color}

The spectral shape of the synchrotron radiation and the projected linear sizes of the synchrotron-radiating regions offer implications for the evolution stage and environment of an RLAGN \citep{miley_distant_2008,murgia_dying_2011,gasperin_radio_2018,hardcastle_radio_2020,zhong_revisiting_2023}.
\apjsa{To characterize the spectral shapes of our UNIONS-VLASS sources, we show the radio color-color diagram between $\tnrd{\alpha}{FIRST=1400MHz}{LoTSS=144Mhz}$ and $\tnrd{\alpha}{VLASS=3000MHz}{FIRST=1400Mhz}$ in Figure~\ref{fig: radio cc}, which includes 30,091 sources detected in VLASS, FIRST, and LoTSS with $\mathtt{alpha\_\{frequency1\}\_\{frequency2\}\_err}<0.2$.}
The expected spectral shapes covering 144~MHz to 3~GHz are indicated by the orange lines and can be divided into four quadrants (labeled Q1--Q4).

Sources located in quadrant 1 have an upturned spectrum, which could be explained by a burst of synchrotron radiation subsequent to the FIRST observations conducted mostly in the 1990s, or the inhomogeneity of the radio regions observed \citep{condon_essential_2016}.
Sources located in quadrant 2 have an inverted spectrum.
One possible explanation is the sudden increase in the radio brightness. 
More commonly, it is due to the fact that synchrotron self-absorption \citep[SSA; e.g.,][]{tingay_investigation_2003,nyland_quasars_2020} dominates synchrotron radiation.
Sources located in quadrant 3 have the most commonly observed power-law distribution of the nonthermal bremsstrahlung.
These sources may have a steepening of the spectral index, forming the curved power-law spectrum as a result of synchrotron losses and/or inverse-Compton losses \citep{dermer_high_2009,blandford_relativistic_2019}.
Sources located in quadrant 4 have a peaked frequency ($\tnr{\nu}{p}$) inasmuch an equilibrium between the brightness and kinetic temperatures of the synchrotron-radiating electrons \citep{longair_high_2011}.
The corresponding radio galaxies could be GHz-peaked ($0.5\leq\tnr{\nu}{p}\leq5$ GHz) or high-frequency peaked ($\tnr{\nu}{p}>5$ GHz) sources depending on their redshifts, which comprises the so-called Peaked-Spectrum (PS) sources that are young radio galaxies (see \citealt{odea_compact_2021} and references therein).
\apjsa{Sources with inverted spectrum between 144~MHz and 3~GHz (in quadrant 2) may also have peaked/turnover frequencies above 3~GHz, making them PS sources.}

\begin{figure*}[tp!]
\includegraphics[width=\textwidth]{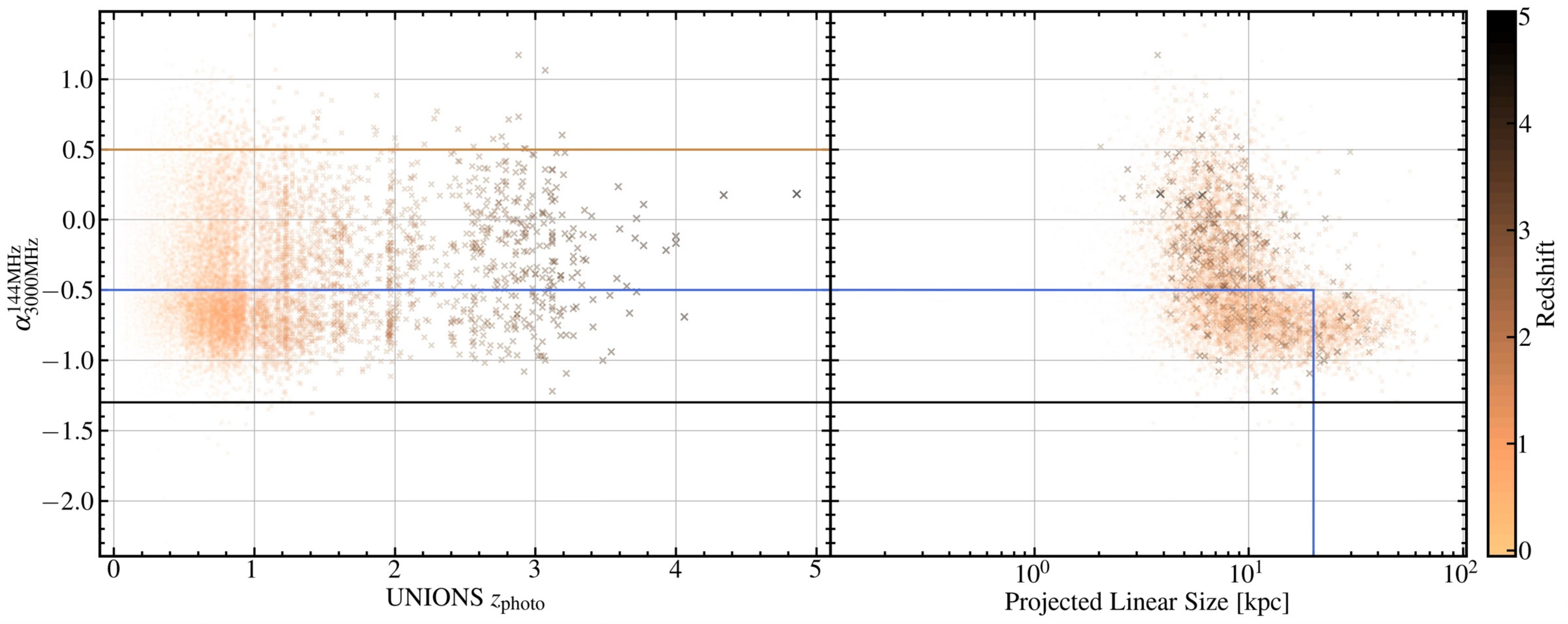}
\caption{Spectral index between LoTSS 144~MHz and VLASS 3~GHz as function of the UNIONS BPZ photo-$z$ and the projected linear size considering the deconvolution of the beam size based on VLASS observations for \apjsa{16,078} UNIONS-VLASS sources reliable in $\tnr{z}{photo}$ and detected in LoTSS.
The orange line indicates $\tnrd{\alpha}{3000MHz}{144MHz}=0.5$.
The black line indicates $\tnrd{\alpha}{3000MHz}{144MHz}=-1.3$, below which the sources are classified as USS sources.
The blue line indicates $\tnrd{\alpha}{3000MHz}{144MHz}=-0.5$ and projected linear size smaller than 20~kpc, and the enclosed region defines CSS sources.
\label{fig: alpha_lv}}
\end{figure*}

\begin{figure*}[tp!]
\includegraphics[width=\textwidth]{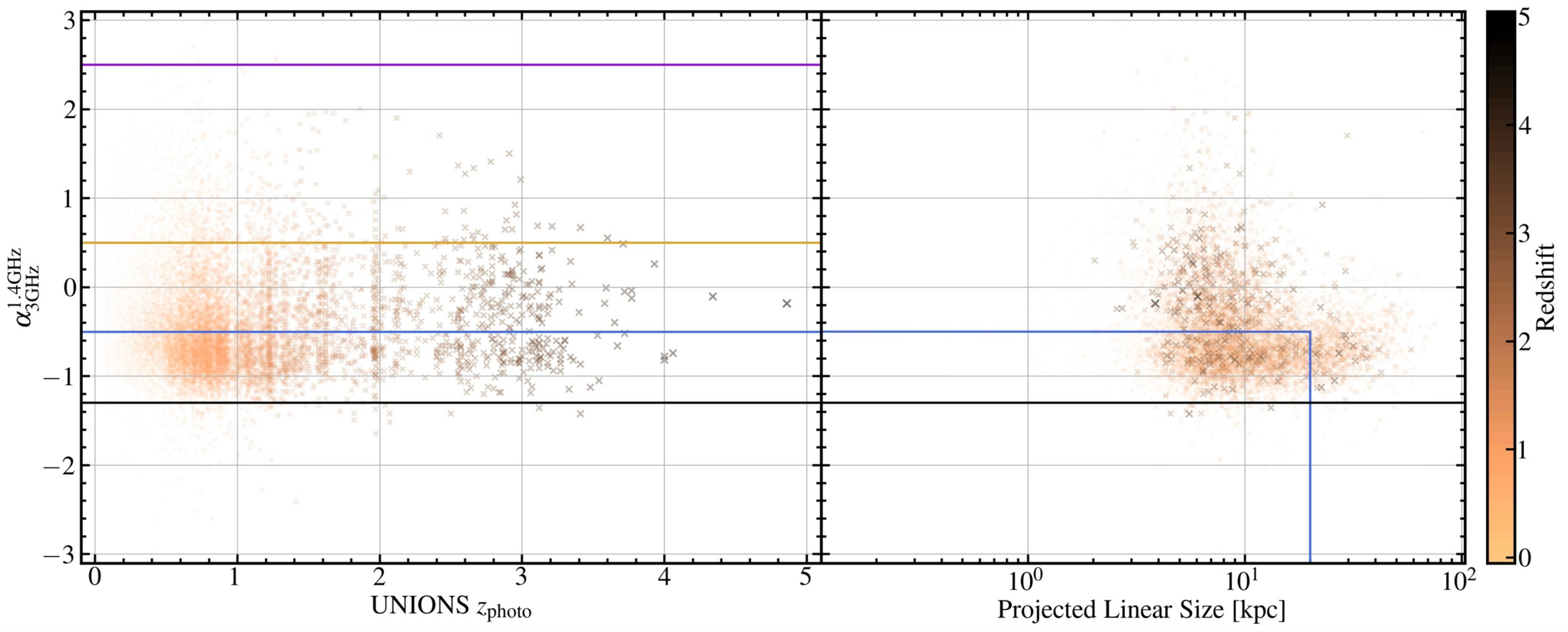}
\caption{Spectral index between FIRST 1.4~GHz and VLASS 3~GHz as function of the UNIONS BPZ photo-$z$ and the projected linear size considering the deconvolution of the beam size based on VLASS observations for $\apjsa{15,906}$ UNIONS-VLASS sources detected in FIRST and are valid in $\tnr{z}{photo}$.
The colored lines are the same as Figure~\ref{fig: alpha_lv}, and an additional violet line indicates the SSA-dominated inverted spectral index of $+2.5$.
\label{fig: alpha_fv}}
\end{figure*}

The sizes of the synchrotron-emitting volume matter when it comes to the classification of radio galaxies.
We further show the spectral index $\tnrd{\alpha}{VLASS=3000MHz}{LoTSS=144MHz}$ ($\apjsa{16,078}$ UNIONS-VLASS-LoTSS sources) and $\tnrd{\alpha}{VLASS=3000MHz}{FIRST=1400MHz}$ ($\apjsa{15,906}$ UNIONS-VLASS-FIRST sources) as a function of photo-$z$ and the VLASS projected linear size of the radio regions deconvolved from the beam size in Figure~\ref{fig: alpha_lv} and \ref{fig: alpha_fv}, respectively, considering only those with reliable spectral index estimates (uncertainties smaller than 0.2).
The black solid line indicates a spectral index of $-1.3$, and sources with a spectral slope steeper than this value are called ultra-steep spectrum (USS; \citealt{de_breuck_sample_2000}).
This could be due to a significant synchrotron aging, suggestive of well-evolved RLAGNs \citep{van_breugel_optical_1984,turner_duty-cycle_2018}.
USS is often used to select high-$z$ RLAGNs as energy losses could be dominated by inverse-Compton scattering, as well as being attributed to the redshifting of the frequency \citep{carilli_radio--submillimeter_1999, morabito_investigating_2018}.
The blue solid line indicates a spectral index of $-0.5$.
If a source has $\alpha<-0.5$ at GHz regime and its projected linear size is more compact than 20~kpc, it is then classified as compact steep spectrum (CSS; \citealt{callingham_extragalactic_2017,nyland_quasars_2020,orienti_young_2023}).
CSS sources are often thought to be RLAGNs at an early evolution stage, otherwise, they are actually evolved populations while their jets are confined by the ISM of their host galaxies \citep{murgia_spectral_2003,mukherjee_relativistic_2016}.
Some sources lie above $\tnrd{\alpha}{3GHz}{1.4GHz}=2.5$, which is beyond the inverted spectral slope expected to be under the dominance of SSA.
They are likely to be transients that undergo a sudden outburst of radio jets.

In this work, we only briefly report the spectral index distribution of our UNIONS-VLASS sources.
A detailed and thorough investigation of the radio classifications of our sources will be presented in the future work (Zhong et al. 2025 in preparation).

\begin{figure}[tp!]
\includegraphics[width=\columnwidth]{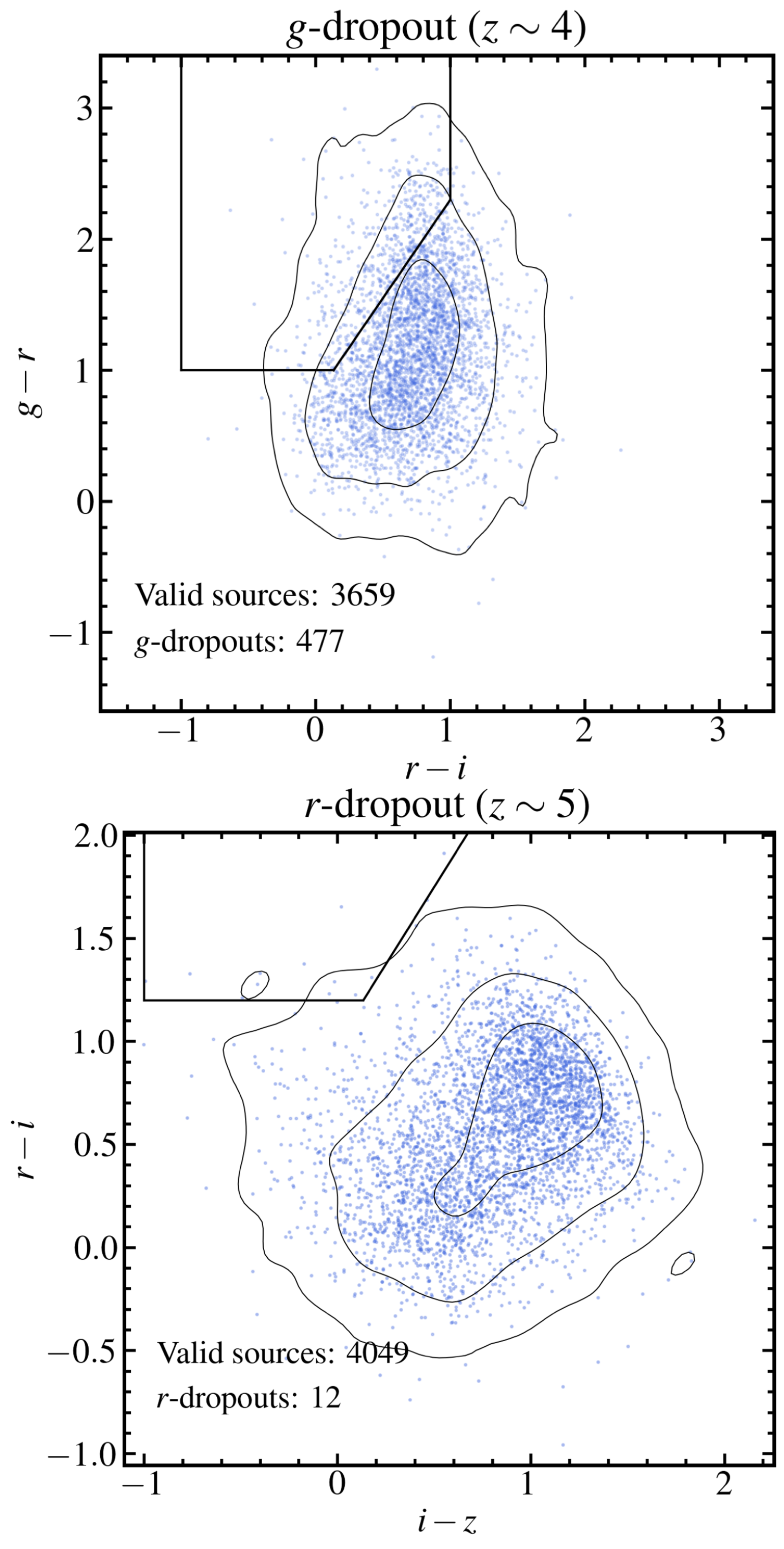}
\caption{Color-color diagrams to select dropout sources.
The upper and lower panels show selections for \textit{g-} ($3<z<5$) and \textit{r-}dropout ($4<z<6$) sources, respectively.
The black lines represent the criteria adopted to select dropouts (Eqs.~(5)-(6)).
The blue dots are all UNIONS-VLASS sources with valid photometry in the corresponding two-color diagram.
The contours are in $1\sigma$, $2\sigma$, and $3\sigma$ confidence levels.
\label{fig: dropouts}}
\end{figure}


\subsection{High-\texorpdfstring{$z$}{TEXT} Candidates with \texorpdfstring{$z\geq3$}{TEXT}}\label{sec: dropouts}
As introduced in \S\,\ref{subsubsec:radio color-color}, the USS characterized by $\alpha<-1.3$ is an efficient technique to select HzRGs.
However, in addition to these USS sources, many HzRGs have been discovered with relatively flat spectral index (e.g., $\tnrd{\alpha}{1.4GHz}{325MHz}=-0.75$ for J163912.11+405236.5 at $z=4.88$, \citealt{jarvis_discovery_2009}; $\tnrd{\alpha}{1.4GHz}{325MHz}=-1.1$ for HSC J083913.17+011308.1 at $z=4.72$, \citealt{yamashita_wide_2020}; see also \citealt{miley_distant_2008}).
As a consequence, a sole reliance on USS could lead to a bias in the statistical physical properties of the HzRGs since the diverse nature could be over-represented by USS.

An alternative approach to select HzRGs, which can circumvent the selection bias arising from the uncertain real spectral index distributions, is the Lyman break technique commonly adopted in the search for high-$z$ galaxies.
The effectiveness of this technique in discovering HzRGs has been proven by \citet{capetti_quest_2025}, who identified 39 $g-$dropout HzRGs expected to be at $3<z<4.5$ over an area of $\sim2,000\sqdeg$, and by \citet{yamamoto_wide_2025}, who constructed a photometric sample of 146 HzRGs at $z\sim4$ within a total survey area of $\sim560\sqdeg$.

The Lyman break color selection technique selects sources that exhibit a clear Lyman break and blue UV continuum in their optical SEDs covered by broadband (\textit{ugriz}) photometry.
Following \citet{hildebrandt_cars_2009} that have used a similar filter set, \textit{g-}dropouts ($3<z<5$) are selected according to
\begin{equation}
    \begin{aligned}
        & g-r>1.0, \\
        & r-i<1.0, \\
        & g-r>1.5(r-i)+0.8,
    \end{aligned}
\end{equation}
and \textit{r-}dropouts ($4<z<6$) are extracted adopting 
\begin{equation}
    \begin{aligned}
        & r-i>1.2, \\
        & i-z<0.7, \\
        & r-i>1.5(i-z)+1.0.
    \end{aligned}
\end{equation}
In each dropout selection, three adjacent bands have to be used.
We therefore only select dropout candidates if the photometric measurements of these bands are not flagged and have valid AB magnitudes (see \S\,\ref{sec: optical-radio correspondence}).
Additionally, for \textit{g-}dropouts, the observation can barely detect fluxes prior to the Lyman break, thus the \textit{u-}band should be fainter than the detection limit with $\mathtt{MAG\_GAAP\_u==99}$, that is, there is no successful photometric measurement of the corresponding source.
Likely, this criterion of \textit{u-}band for \textit{g-}dropouts should be applied to both \textit{u-} and \textit{g-}bands for \textit{r-}dropouts.
Moreover, to ensure these high-$z$ candidates are real detections, we further require their $\tnr{i}{ABmag}$ to be brighter than the $10\sigma$ detection limit 23.8.
Furthermore, to avoid foreground contaminants, we apply a magnitude cut of $\tnr{i}{ABmag}\geq23$.

We show \textit{g-} and \textit{r-}dropout candidates in the upper and lower panels of Figure~\ref{fig: dropouts}, respectively.
The above criteria produce 4,049 valid sources for the \textit{r-}dropouts, and only 12 of them can be at $z\sim5$, which is reasonable considering the limited depth of UNIONS.
Of $\upda{63,019}$ UNIONS-VLASS sources with reliable BPZ photo-$z$, there are only 2 at $z>4.5$, comparable with the \textit{r-}dropout selections.

The number of \textit{g-}dropout-selected sources is $477$, which exceeds the number counts of \upda{53} expected from reliable $3.5<\tnr{z}{photo}<4.5$ estimates by almost one order of magnitude.
This potential excess in the number counts of \textit{g-}dropouts -- if they are not bona fide sources at $z\sim4$ -- could be explained by the contamination of foreground sources such as red elliptical galaxies at intermediate redshifts, whose $\tnr{D}{4000}$ break due to the absorption of old stellar populations mimic the Lyman break feature.
The photometric errors of these foreground contaminations satisfy the selection criteria of the \textit{g-}dropout \citep[e.g.,][]{ono_great_2018}.
For Subaru HSC, which has comparable optical number counts and similar efficient broadband wavelengths to those of UNIONS, \citet{ono_great_2018} found the contamination rate in the $i$-band magnitude bin of 22.5 can reach $\sim68\%$ and decreases to $\sim41\%$ at $\tnr{i}{ABmag}=23.1$ and to $\sim35\%$ at $\tnr{i}{ABmag}=23.7$ for \textit{g-}dropouts.
Therefore, even if we only select UNIONS galaxies at the faint end of $i$-band ($\tnr{i}{AB}>23$), the contamination problem can still persist.
\apjsb{Similarly, for $r$-dropouts, \citet{ono_great_2018} reported contaminate rates of approximately $59\%$ at $\tnr{z}{ABmag}=22.3$ and around $23\%$ at $\tnr{z}{ABmag}=23.5$, implying that merely $<5$ $r$-dropouts could be true positives if applying these estimates to UNIONS-VLASS.}
Nonetheless, under the optimistic assumption of HSC-level contamination rates, our UNIONS-VLASS radio galaxy catalog is expected to yield more than 200 HzRGs at $z \sim 4$, offering strong potential to significantly enhance the statistical robustness of studies on HzRGs.

All in all, these expectations for high-\textit{z} RLAGNs are subject to the reliability of the photometry of UNIONS.
A deeper investigation into the HzRGs will be conducted in future work using spectral energy distribution fittings combining IR data (Zhong et al. in preparation).

\subsection{Caveat to the Catalog}
\label{sec:caveat}
\apjsa{
The effect of applying $\mathtt{NN\_dist>30''}$ to VLASS2 is clearly reflected in the projected linear size, that is, the majority of our RGs are compact ($\tnr{l}{s}\leq20$~kpc), which means our UNIONS-VLASS RGs have the tendency to be young and evolving populations.
Therefore, for HzRGs that are often selected by the USS feature in previous studies, there are only a few of them showing USS at $z>2$ amongst $15,906$ UNIONS-VLASS-FIRST sources with valid $\tnrd{\alpha}{VLASS}{FIRST}$.
In addition to the $\mathtt{NN\_dist}$ cut, this global trend of being compact for our RGs is more a result of the limited search radius of $1.5''$, as shown in Figure~\ref{fig: nn_sep} and discussed in \S\,\ref{sec: nn_dist cut}.
To maintain the current cleanliness of the UNIONS-VLASS catalog, if $\mathtt{NN\_dist\leq30''}$ is not adopted, the inclusion of bright and extended radio sources in the UNIONS-VLASS leads to an increase of 20,000 in the sample size, which is about 1/3 of the total number of sources with $\mathtt{NN\_dist\leq30''}$ in VLASS2.
We, thereby, remind the readers to be cautious about the intrinsic and unresolved bias arising from the design of this work.
Nonetheless, the vast size of this catalog still ensures statistical studies of radio AGNs across various evolution stages, offering valuable insights into the connections between radio AGNs and their host galaxies.
}

\section{Summary}\label{sec: summary}
We presented the results of searching for the optical counterparts of radio sources using the grand UNIONS catalog and the VLASS catalog.
This marks the first step of the ongoing project, UNIONS-VLASS radio galaxies in the Euclid wide survey area (UNVEIL).
The current angular separation-based cross-match between UNIONS and VLASS found 146,212 radio galaxies/quasars spanning a wide area of $\sim4,200\ \mathrm{deg^2}$ in the northern and southern Galactic caps.
The fraction of matches reaches 0.594, beyond the previous FIRST 1.4~GHz radio galaxy catalog based on the SDSS and Subaru HSC catalogs.

Setting UNIONS-VLASS as the reference catalog, we further performed cross-match with FIRST 1.4~GHz and LoTSS 144~MHz catalogs, yielding 79,638 UNIONS-VLASS-FIRST, 101,671 UNIONS-VLASS-LoTSS, and 64,672 UNIONS-VLASS-FIRST-LoTSS radio galaxies, respectively.
We identified radio-loud AGNs based on the observed radio loudness ($\robs$) defined as the ratio of 1.4~GHz radio to \textit{r-}band optical flux densities.
Of 146,212 ``clean'' UNIONS-VLASS radio galaxies, 138,266 ($\sim94.6\%$) are RL and 7,496 ($\sim5.4\%$) are RQ populations.
Alternatively, RLAGNs can be selected by their 1.4 spectral luminosities $\tnr{L}{1.4GHz}>10^{24}\whz$.
Under this condition, at $z\gtrsim1$, both VLASS and FIRST cannot detect any RQ populations.
RLAGNs selections based on $\robs$ and $\tnr{L}{1.4GHz}$ are common for 49,012 out of 63,019 sources with valid UNIONS photometric redshift estimates.

We presented radio color-color ($\tnrd{\alpha}{1400MHz}{144MHz}-\tnrd{\alpha}{3000MHz}{1400MHz}$) based on LoTSS, FIRST, and VLASS observations, and further plotted spectral indices as a function of redshift and the projected linear size of the synchrotron-emitting volume.
We found a great diversity of radio classifications, including compact steep spectrum and peaked spectrum sources that are likely young radio galaxies, ultra steep spectrum sources which may have significantly evolved, and inverted and upturned spectrum that may be linked to transient sources.
There are $\sim10,000$ additional radio galaxies corresponding to more than one optical counterpart.
These sources are merger candidates that host RLAGNs.
The radio classifications/RLAGN evolution and merger-RLAGN connections will be further studied in future works.

The UNIONS internal data release used for this work does not cover the full UNIONS footprint in both NGC and SGC.
In addition, UNIONS is an ongoing survey that will eventually cover the entire northern hemisphere.
Moreover, the VLASS Epoch 3 observation has just been completed, and the concatenated catalog reaching a great depth has not been released.
Therefore, our UNIONS-VLASS catalog has great potential for expansion.
The full UNIONS footprint will be covered by the \textit{Euclid} wide survey, and this work serves as a valuable pavement for the future \textit{Euclid} radio galaxy catalog.
Our UNIONS-VLASS radio galaxy catalog will be observed in the Subaru PFS filler program.
By virtue of the broad spectroscopic coverage ($\sim3000-20000\AA$) and imaging data, the abundance of RLAGNs in the UNIONS-VLASS catalog will profoundly facilitate multifaceted statistical studies of radio galaxies and beyond.

\section*{Acknowledgments}
We wish to thank Takuji Yamashita and Hongming Tang for the discussions.
YZ is supported by Japan Society for the Promotion of Science Research Fellowship for Young Scientists.
This work is supported by Japan Society for the Promotion of Science KAKENHI (23H00131; Akio~K.~Inoue, 25K01043; K.~Ichikawa, 24K22894; M.~Onoue).

We are honored and grateful for the opportunity of observing the Universe from Maunakea and Haleakala, which both have cultural, historical and natural significance in Hawaii. 
This work is based on data obtained as part of the Canada-France Imaging Survey, a CFHT large program of the National Research Council of Canada and the French Centre National de la Recherche Scientifique. Based on observations obtained with MegaPrime/MegaCam, a joint project of CFHT and CEA Saclay, at the Canada-France-Hawaii Telescope (CFHT) which is operated by the National Research Council (NRC) of Canada, the Institut National des Science de l’Univers (INSU) of the Centre National de la Recherche Scientifique (CNRS) of France, and the University of Hawaii. 
This research used the facilities of the Canadian Astronomy Data Centre operated by the National Research Council of Canada with the support of the Canadian Space Agency.
This research is based in part on data collected at Subaru Telescope, which is operated by the National Astronomical Observatory of Japan.
Pan-STARRS is a project of the Institute for Astronomy of the University of Hawaii, and is supported by the NASA SSO Near Earth Observation Program under grants 80NSSC18K0971, NNX14AM74G, NNX12AR65G, NNX13AQ47G, NNX08AR22G, 80NSSC21K1572 and by the State of Hawaii.

The National Radio Astronomy Observatory is a facility of the National Science Foundation operated under cooperative agreement by Associated Universities, Inc.
CIRADA is a partnership between the University of Toronto, the University of Alberta, the University of Manitoba, the University of British Columbia, McGill University and Queen's University, in collaboration with National Research Council of Canada, the US National Radio Astronomy Observatory, and Australia's Commonwealth Scientific and Industrial Research Organization. CIRADA is funded by a grant from the Canadian Foundation for Innovation 2017 Innovation Fund (Project 35999) and the by the Provinces of Ontario, British Columbia, Alberta, Manitoba and Quebec.

This research used the facilities of the Canadian Astronomy Data Centre operated by the National Research Council of Canada with the support of the Canadian Space Agency.

%

\vspace{5mm}
\facilities{VLA, CFHT, Pan-STARRS, Subaru HSC, and LOFAR}


\software{astropy \citep{astropy_collaboration_astropy_2013,astropy_collaboration_astropy_2018}; astroquery \citep{ginsburg_astroquery_2019}
          }



\appendix
\section{Radio Morphology of Sources with $\mathtt{NN\_dist}<30''$}
\begin{figure*}[tp!]
\includegraphics[width=\columnwidth]{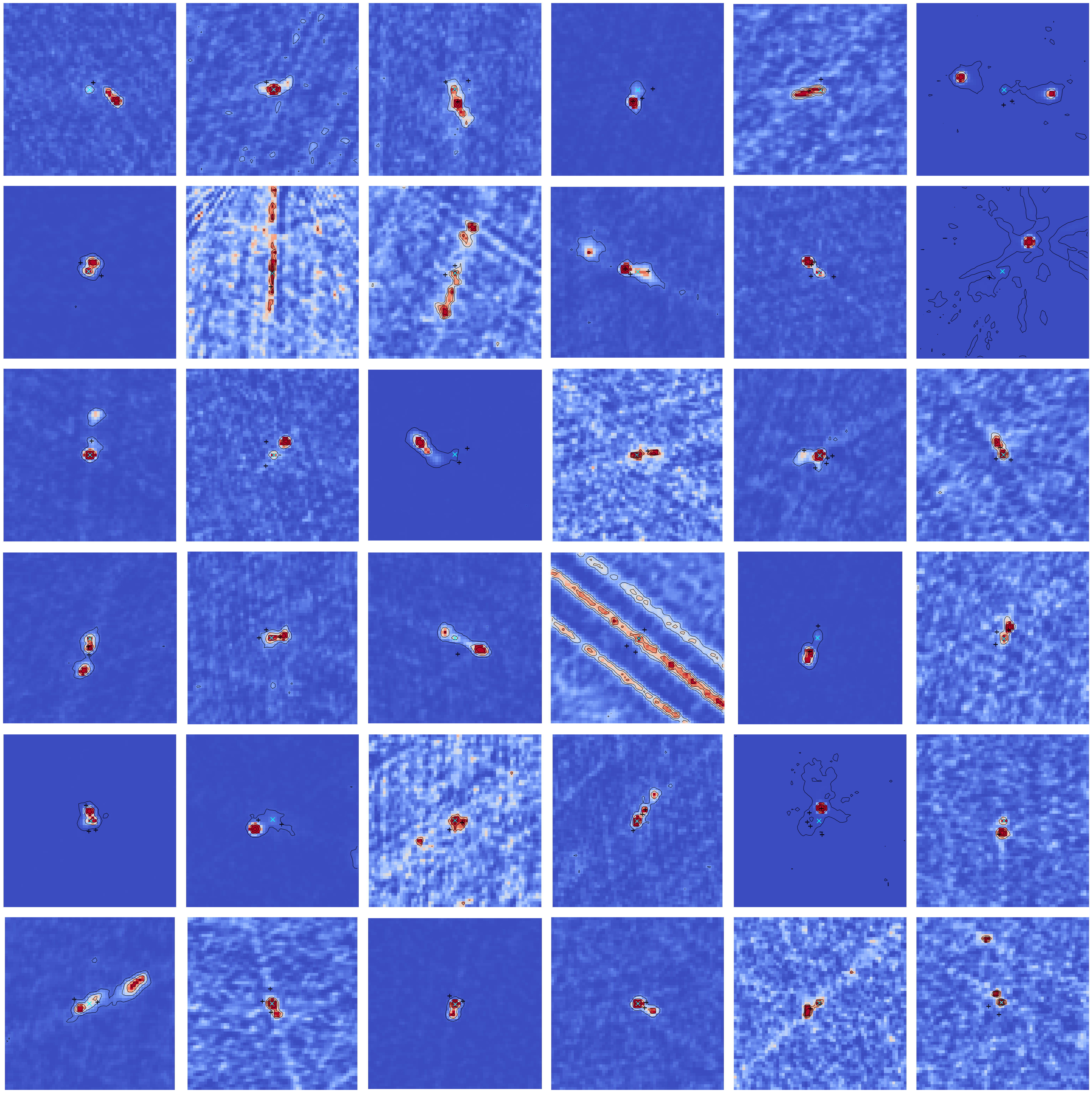}
\caption{VLASS Epoch 2 QL images of VLASS sources with $\mathtt{NN\_dist}<30''$.
Cyan crosses represent the centroids of VLASS sources and black pluses represent the centroids of UNIONS sources cross-matched within a search radius of $2.5''$.
\label{fig: vlass nn imaging}}
\end{figure*}

We randomly choose 36 VLASS radio sources with $\mathtt{NN\_dist}<30''$ and show their Epoch 2 QL images in Figure~\ref{fig: vlass nn imaging}.

\section{Additional Astrometric Comparisons}
\begin{figure}[tp!]
\includegraphics[width=\columnwidth]{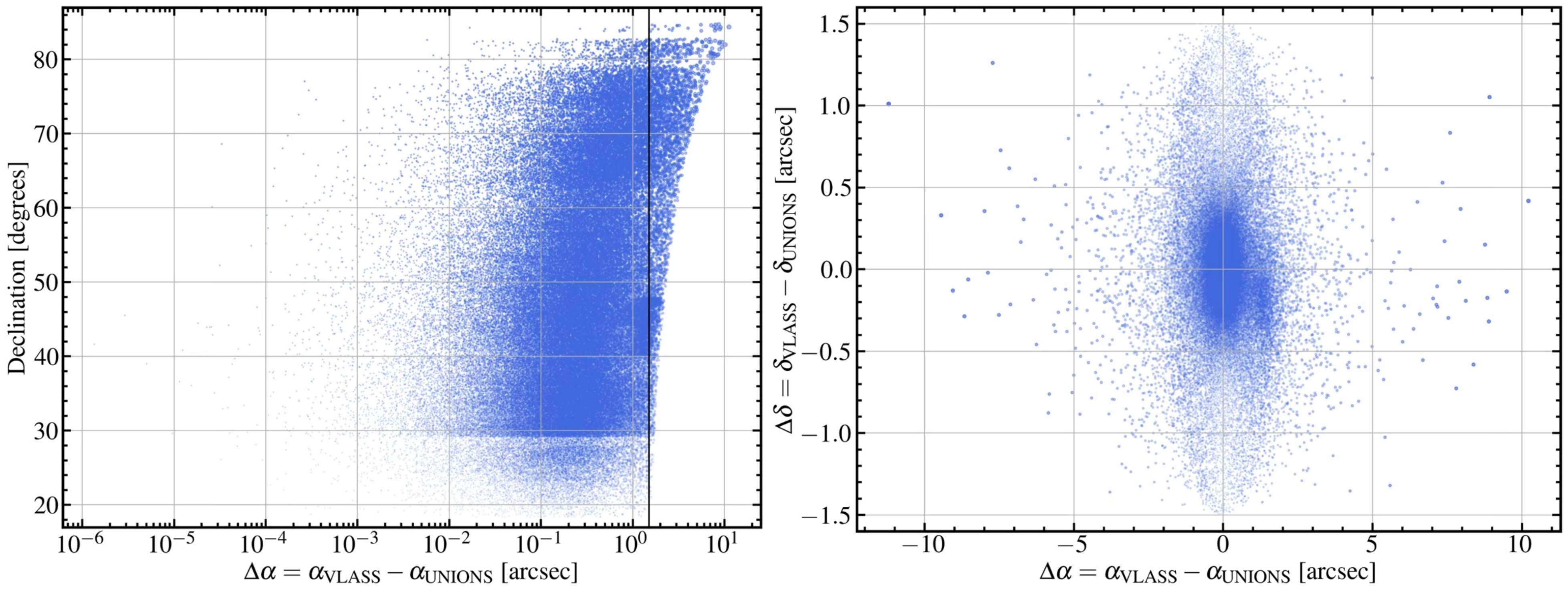}
\caption{\textit{Left}: the offsets along the R.A. as a function of the declination for 146,212 ``clean'' UNIONS-VLASS radio galaxies, where the vertical line indicates the angular separation-based cross-match radius of $1.5\arcsec$.
The offset along R.A. suffers from a more severe projection effect at higher declinations.
\textit{Right}: the plane of offset along R.A. versus the offset along declination.
\label{fig: astrometry app}}
\end{figure}

We show the absolute values of astrometric offsets along R.A. for 146,212 UNIONS-VLASS radio galaxies cross-matched via angular separations $\leq1.5\arcsec$ between UNIONS and VLASS sources as a function of the declination in the left panel of Figure~\ref{fig: astrometry app}.
It makes clear that large astrometric offsets along R.A. are attributed to the projection effect, and this worsens as the declination increases.
For previous studies based on VLA FIRST and adopting positional cross-match, this situation is barely considered since the sky coverage of FIRST is limited to $\mathrm{DEC<65^\circ}$, and the projection effect does not lead to a significant excess of the astrometric offset along R.A.
However, for VLASS that approaches the north pole, an angular separation-based cross-match is preferred over the positional one, otherwise, a small fraction of the radio sources can be missed in the UNIONS-VLASS ($\sim4\%$), as well as in other future catalogs using VLASS.

\section{UNIONS-VLASS \textit{r-}band Source Counts for RL and RQAGNs}

\begin{figure}[tp!]
\includegraphics[width=\columnwidth]{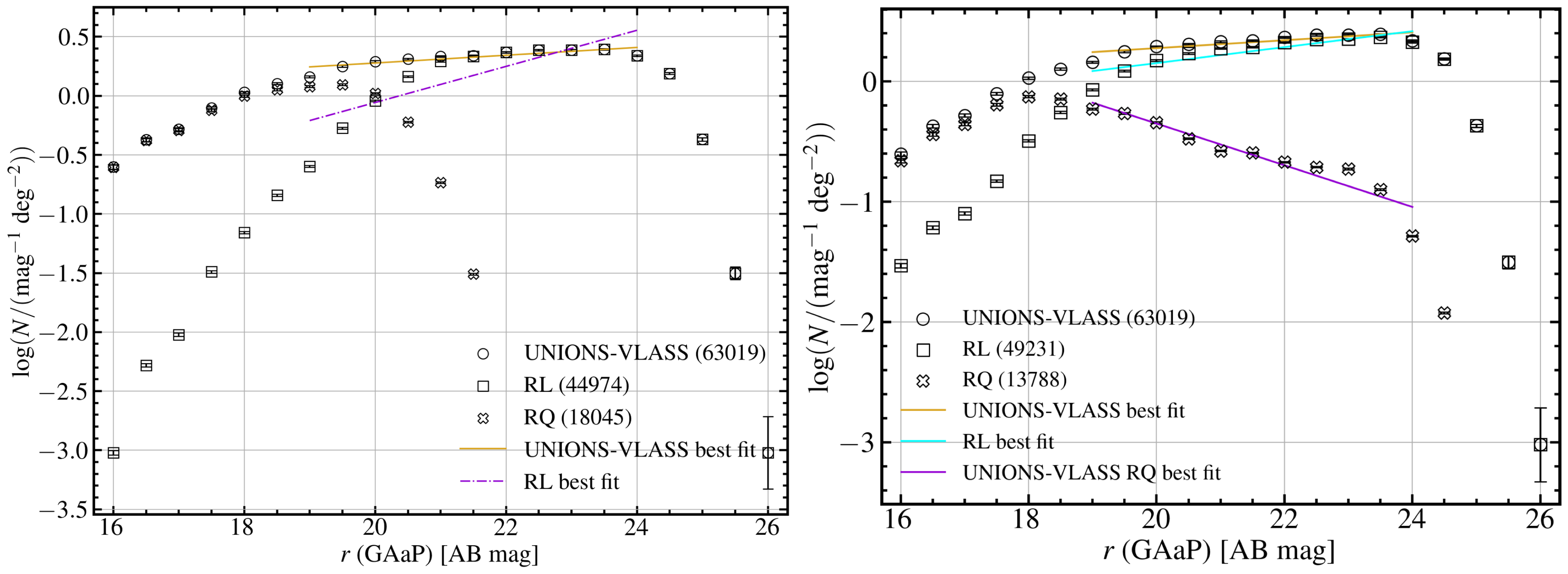}
\caption{Optical $r$-band source counts for \upda{63,019} UNIONS-VLASS radio galaxies (circles).
In both panels, squares represent RL and crosses represent RQ populations.
\textit{Left}: the UNIONS-VLASS sources are divided into RL and RQAGNs based on the radio loudness, where RLAGNs have $\log\robs>1$.
\textit{Right}: RL and RQAGNs are separated via the 1.4~GHz spectral luminosity, where RLAGNs have $\tnr{L}{1.4GHz}>10^{24}\whz$.
\label{fig: optical number counts r rlrq}}
\end{figure}

We divide \upda{63,019} UNIONS-VLASS sources with valid photo-$z$ into RL (squares) and RQ (crosses) subsample based on the observed radio loudness ($\robs$; see \S\,\ref{sec: radio loudness} for details) and show the optical $r$-band source counts for the two populations, respectively, in the left panel of Figure~\ref{fig: optical number counts r rlrq}.
We see that RLAGNs dominate the AGN populations whose optical counterparts lie above $r\gtrsim20$.
On the other hand, the brightest optical counterparts are almost host galaxies to RQAGNs, and there are no RQAGNs in optically faint galaxies ($r>20$).
This is consistent with the distribution of the radio loudness against $\tnr{r}{ABmag}$.
For optically luminous sources, their large optical flux densities make it more difficult for them be radio-loud even though they are powerful in synchrotron radiation.
In this manner, RLAGNs are preferentially associated with massive systems with inactive star-forming actives, while RQAGNs are more likely to reside in systems with intense star formation. 

We further divide RL and RQ populations based on their 1.4~GHz spectral luminosities (see \S\,\ref{sec: radio loudness} for details).
We see that RQAGNs are still the dominant populations for galaxies at $\tnr{r}{ABmag}<18.5$.
However, at $\tnr{r}{ABmag}>20$, the source counts of RQAGNs do not decrease as steeply as the trend based on $\robs$, which is similar to that of radio quasars (see, e.g, \citealt{yamashita_wide_2018}).
With the increasing $r$-band magnitude, there is a lower possibility of finding RQAGNs compared to RL populations.
In summary, from this aspect, both RL and RQAGNs are hosted by a large variety of galaxy populations.

\section{Radio-loud AGNs}
Star formation can contribute to the non-thermal emission, and rigorously $\log\robs>2$ may be adopted to exclude star-forming/starburst galaxies that mimic RLAGNs \citep[e.g.,][]{chiaberge_origin_2011}.
This reduces our RLAGNs to 109,901 ($\sim77.0\%$).
The spectral luminosity of free-free emission attributed to star formation can be quantified by \citep{condon_radio_1992}:
\begin{equation}
    \lrp{\frac{\tnr{L}{T}}{\whz}}\sim5.5\times10^{20}\lrp{\frac{\nu}{\mathrm{GHz}}}^{-0.1}\left[\frac{\mathrm{SFR(\mathit{\tnr{M}{star}}\geq5\msun)}}{\sfr}\right],
\end{equation}
and the non-thermal emission related to SF activities follows 
\begin{equation}
    \lrp{\frac{\tnr{L}{NT}}{\whz}}\sim5.3\times10^{21}\lrp{\frac{\nu}{\mathrm{GHz}}}^{-\alpha}\left[\frac{\mathrm{SFR(\mathit{\tnr{M}{star}}\geq5\msun)}}{\sfr}\right].
\end{equation}
Assuming an upper limit of $\mathrm{SFR}=300\ \sfr$ for star-forming main sequence galaxies at $z\gtrsim2$ \citep{popesso_main_2023}, the corresponding luminosities are $\tnr{L}{T, SF}\sim1.5\times10^{23}\whz$ and $\tnr{L}{NT, SF}\sim1.3\times10^{24}\whz$.
In Figure~\ref{fig: radio luminosity} for $\tnr{L}{1.4GHz}$ and $\tnr{L}{3GHz}$ as a function of $z$, this $\tnr{L}{NT,SF}$ only selects radio galaxies at $z\lesssim1$ where the host galaxies should have much smaller values of SFR compared to high-\textit{z} sources even if they are rarely observed starburst systems.
The real contributions of star-forming activities to the observed synchrotron radiation should be insignificant ($\lesssim1.5\times10^{22}\whz$ assuming $\mathrm{SFR}=40\ \sfr$ at $z\sim0.5$).
Therefore, $\log\robs>1$ can be safely adopted to select RLAGNs for the UNIONS-VLASS sample.

We see correlations between $\tnr{L}{1.4GHz}-\robs$ and $\robs-\tnr{z}{photo}$, and these correlations are likely a propagation of the correlation between $\tnr{i}{ABmag}-\tnr{z}{photo}$ shown in the left panel of Figure~\ref{fig: imag, radio mag} that reflects the cosmological surface brightness dimming.
As a result, high-$z$ sources have higher possibilities of being classified as RLAGNs because of their optical faintness.
In the right panel of Figure~\ref{fig: imag, radio mag}, we plot absolute radio magnitude $M_t$ as a function of the observed radio loudness $\robs$, where $M_t$ is defined as \citep{ivezic_optical_2002}
\begin{equation}
    M_t=-2.5\log\lrp{\frac{\tnr{L}{radio}}{\tnr{L}{AB}}}.
\end{equation}
$\tnr{L}{radio}$ is the specific or spectral radio luminosity, and $\tnr{L}{AB}=4.345\times10^{13}\whz$ is the specific luminosity of a source whose specific flux
is 3631 Jy at a distance of 10 pc.
Normally, $M_t=-25.9$ is chosen as the RL/RQ demarcation, corresponding to $\tnr{L}{1.4GHz}\sim10^{24}\tnrd{h}{50}{-2}\whz$.
For 146,212 UNIONS-VLASS sources with valid \textit{r-}band photometry to calculate $\robs$, the two thresholds to select RLAGNs are the same for 83.3\%.

\begin{figure}[tp!]
\includegraphics[width=\columnwidth]{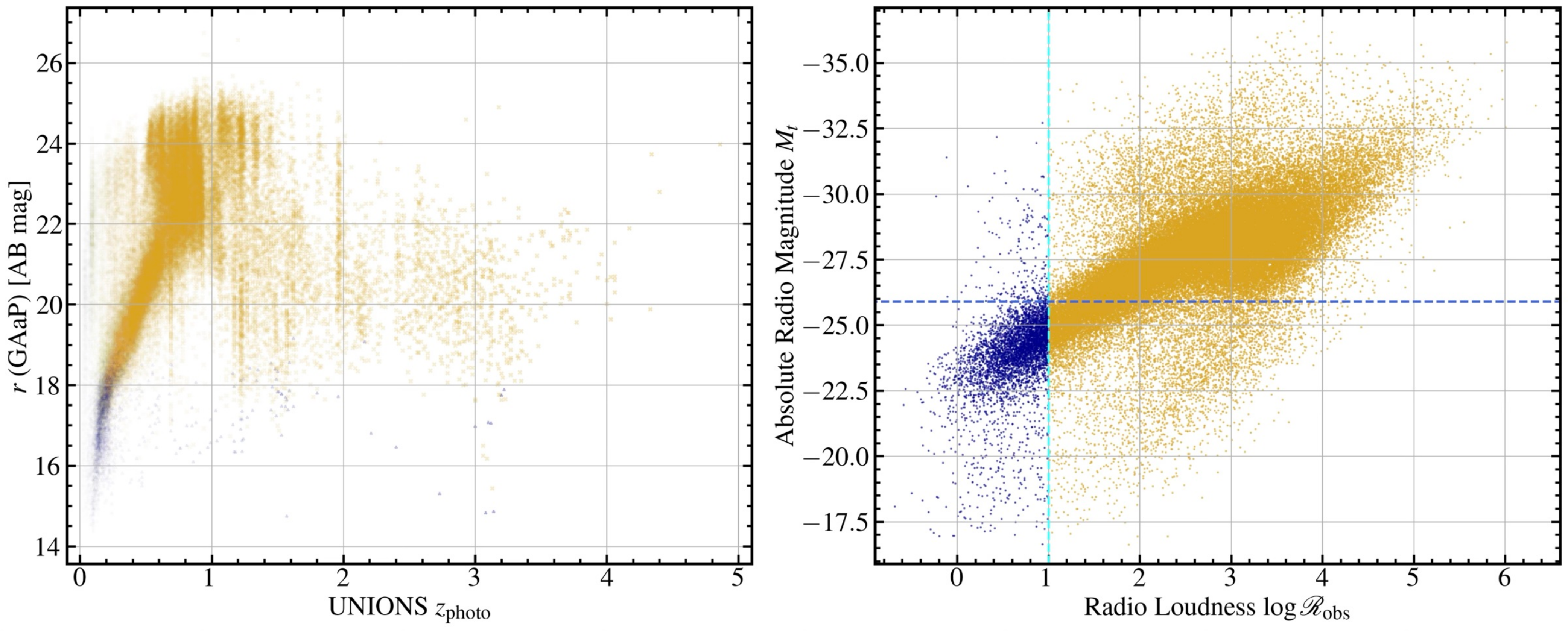}
\caption{\textit{Left}: UNIONS $r$-band photometry as a function of the photometric redshift.
\textit{Right}: Absolute radio magnitude as a function of the observed radio loudness, where the dashed line indicates the RL/RQ demarcation of $M_t=-25.9$.
The golden dots represent RLAGNs based on $\log\robs>1$ and the violet ones are RQAGNs defined by $\log\robs<1$.
\label{fig: imag, radio mag}}
\end{figure}

\section{UNIONS2000 versus UNIONS5000 for \textit{i-}band Photometry}
\begin{figure}[tp!]
\centering
\includegraphics[width=0.5\columnwidth]{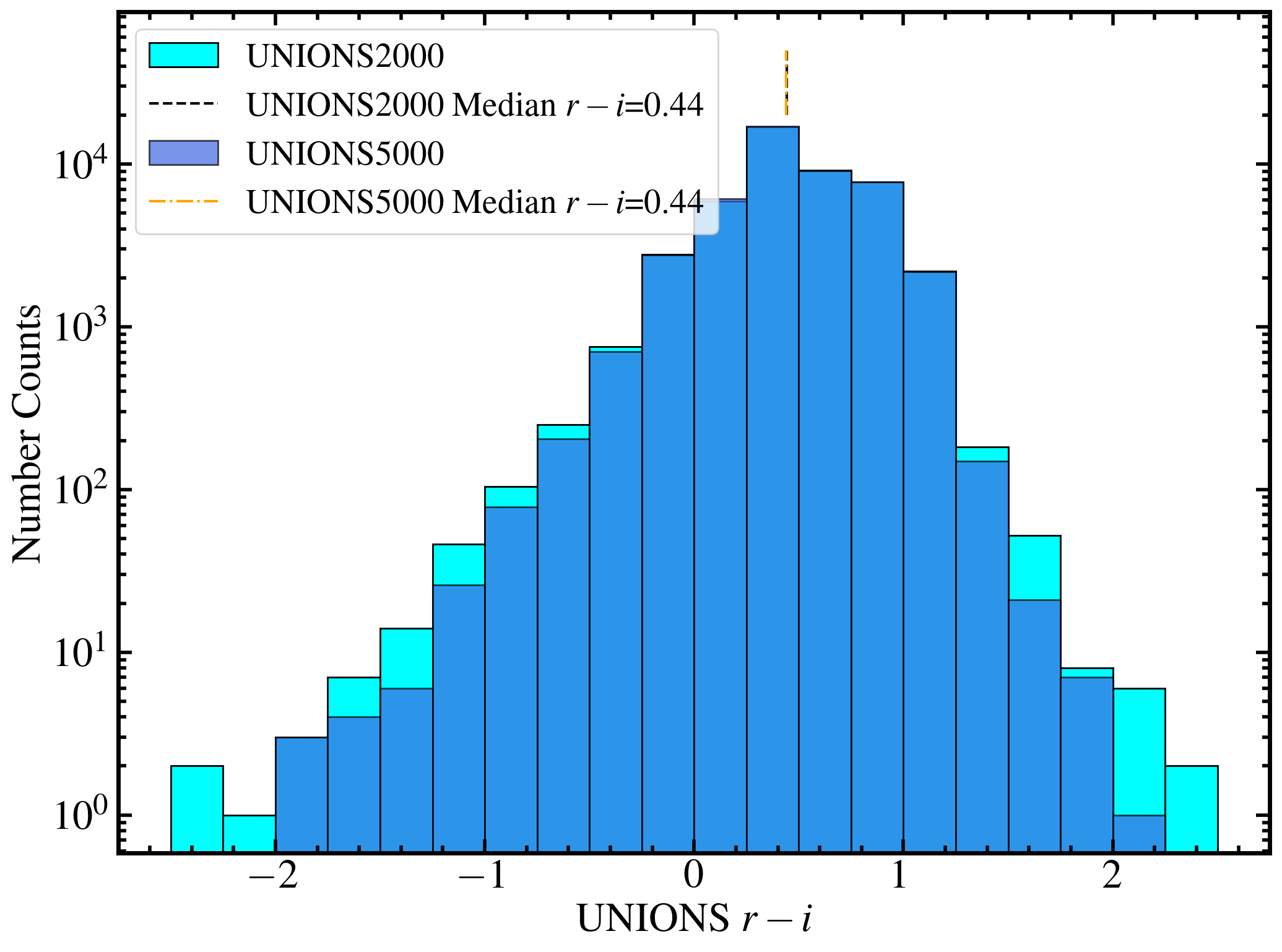}
\caption{The distribution of $\tnr{r}{ABmag}-\tnr{i}{ABmag}$ for radio galaxies commonly in catalogs using UNIONS2000 (cyan) and UNIONS5000 (blue).
\label{fig: unions2000 vs 5000}}
\end{figure}

We compare the distributions of the $r-i$ color between UNIONS2000 and UNIONS5000 for 76,454 UNIONS-VLASS radio galaxies produced in both data releases in Figure~\ref{fig: unions2000 vs 5000}.
The median difference between $\tnr{r}{ABmag}$ and $\tnr{i}{ABmag}$ is 0.44 in both UNIONS2000 and UNIONS5000.
However, in comparison with UNIONS2000, the distribution of $r-i$ in UNIONS5000 is more convergent, which is positively attributed to that the UNIONS5000 has been additionally processed using Pan-STARRS DR3 $i$-band data.


\section{Catalog columns}
We show delivered columns of our UNIONS-VLASS radio galaxy catalog in Table~\ref{tab: catalog columns}, including VLASS whose radio source is set as the reference at 3~GHz, UNIONS (optical; \textit{ugriz}), SDSS (optical; \textit{ugriz}), Pan-STARRS (optical to NIR; \textit{grizy}), AllWISE (NIR to MIR; 3.4, 4.6, 12, and 22 $\micron$m), unWISE (NIR; 3.4 and 4.6 $\micron$m), FIRST (1.4~GHz), and LoTSS (144~MHz), as well as properties calculated in this work.

\begin{longtable}{llll}
\label{tab: catalog columns}\\
\caption{Columns for the Main Catalog} \\
\hline
Column name & Type &  Units &  Description \\
\hline
\endfirsthead

\hline
Column name &  Type &  Units &  Description \\
\hline
\endhead

\hline
\endfoot

\hline
\endlastfoot

\hline
\multicolumn{4}{c}{Columns for VLASS \citep{gordon_quick_2021}} \\
Component\_name & string &  & Unique name of the PyBDSF component\\
RA & double & deg & Right ascension of component [deg]\\
DEC & double & deg & Declination of component [deg]\\
E\_RA & double & deg & Error in RA [deg]\\
E\_DEC & double & deg & Error in dec [deg]\\
Total\_flux & double & mJy & Integrated flux of the component [mJy]\\
E\_Total\_flux & double & mJy & Error in integrated flux [mJy]\\
Peak\_flux & double & mJy/beam & Peak flux of component [mJy/beam]\\
E\_Peak\_flux & double & mJy/beam & Error in component peak flux [mJy/beam]\\
Maj & double & arcsec & Component major axis size (FWHM) [arcsec]\\
E\_Maj & double & arcsec & Error in Maj [arcsec]\\
Min & double & arcsec & Component minor axis size (FWHM) [arcsec]\\
E\_Min & double & arcsec & Error in Min [arcsec]\\
PA & double & deg & Component position angle east of north [deg]\\
E\_PA & double & deg & Error in PA [deg]\\
S\_Code & string &  & PyBDSF source type\\
DC\_Maj & double & arcsec & Deconvolved component major axis size (FWHM) [arcsec]\\
E\_DC\_Maj & double & arcsec & 1 sigma error in DC\_Maj [arcsec]\\
DC\_Min & double & arcsec & Deconvolved component minor axis size (FWHM) [arcsec]\\
E\_DC\_Min & double & arcsec & 1 sigma error in DC\_Min [arcsec]\\
DC\_PA & double & deg & Deconvolved component position angle east of north [deg]\\
E\_DC\_PA & double & deg & 1 sigma error in DC\_PA [deg]\\
Peak\_to\_ring & double &  & Ratio of the Peak\_flux to the maximum flux in annulus of $r=5\arcsec$ to $R = 10\arcsec$ \\
& & & centred on component RA Dec\\
Duplicate\_flag & double &  & Flag to denote the duplicate status of a component\\
Quality\_flag & double &  & QA flag to deal with spurious detections and duplicates due to \\
& & & overlap between VLASS tile edges\\
Source\_name & string &  & Name of the distinct radio source to which this component has been assigned\\
Source\_type & string &  & The type of source the component has been assigned to\\
NN\_dist & double & arcsec & Angular distance to nearest other component in the catalogue [arcsec]\\
BMAJ & double & arcsec & Major axis of the beam [arcsec]\\
BMIN & double & arcsec & Minor axis of the beam [arcsec]\\
BPA & double & deg & Beam position angle east of north [deg]\\
\hline
\multicolumn{4}{c}{Columns for UNIONS \citep{gwyn_unions_2025}} \\
ALPHA\_J2000 & Double & deg &\\
DELTA\_J2000 & Double & deg &\\
MAG\_GAAP\_u & Float & mag & u-band GAaP magnitude optimal min\_aper\\
MAGERR\_GAAP\_u & Float & mag & u-band GAaP magnitude error optimal min\_aper\\
FLAG\_GAAP\_u & Short &  & GAaP Flag for MAG\_GAAP\_u optimal min\_aper\\
MAG\_GAAP\_g & Float & mag & g-band GAaP magnitude optimal min\_aper\\
MAGERR\_GAAP\_g & Float & mag & g-band GAaP magnitude error optimal min\_aper\\
FLAG\_GAAP\_g & Short &  & GAaP Flag for MAG\_GAAP\_g optimal min\_aper\\
MAG\_GAAP\_r & Float & mag & r-band GAaP magnitude optimal min\_aper\\
MAGERR\_GAAP\_r & Float & mag & r-band GAaP magnitude error optimal min\_aper\\
FLAG\_GAAP\_r & Short &  & GAaP Flag for MAG\_GAAP\_r optimal min\_aper\\
MAG\_GAAP\_i & Float & mag & i-band GAaP magnitude optimal min\_aper\\
MAGERR\_GAAP\_i & Float & mag & i-band GAaP magnitude error optimal min\_aper\\
FLAG\_GAAP\_i & Short &  & GAaP Flag for MAG\_GAAP\_i optimal min\_aper\\
MAG\_GAAP\_z & Float & mag & z-band GAaP magnitude optimal min\_aper\\
MAGERR\_GAAP\_z & Float & mag & z-band GAaP magnitude error optimal min\_aper\\
FLAG\_GAAP\_z & Short &  & GAaP Flag for MAG\_GAAP\_z optimal min\_aper\\
Z\_B & Double &  & 9-band BPZ redshift estimate; peak of posterior probability distribu\\
Z\_B\_MIN & Double &  & Lower bound of the 68\% confidence interval of Z\_B\\
Z\_B\_MAX & Double &  & Upper bound of the 68\% confidence interval of Z\_B \\
\hline
\multicolumn{4}{c}{Columns for SDSS DR16 \citep{ahumada_16th_2020}} \\
objID & Long &  & SDSS unique object identifier (1) \\
RA\_ICRS & Double & deg & Right Ascension of the object (ICRS) at Epoch="MJD" (ra) \\
DE\_ICRS & Double & deg & Declination of the object (ICRS) at Epoch="MJD" (dec) \\
mode & Short &  & [1/4] PhotoMode (1=primary, 2=secondary, 3=family, 4=outside) \\
class & String &  & [0/8] Type of object (3=galaxy, 6=star) (type) (2) \\
clean & Short &  & [0/1] Clean photometry flag (1=clean; 0=unclean) \\
e\_RA\_ICRS & Double & arcsec & [0/51703] Mean error on RAdeg (raErr) \\
e\_DE\_ICRS & Double & arcsec & [0/22795] Mean error on DEdeg (decErr) \\
umag & Float & mag & [4/37.3] Model magnitude in u filter AB scale (u) (7) \\
gmag & Float & mag & [3.7/39.6] Model magnitude in g filter AB scale (g) (7) \\
rmag & Float & mag & [3.8/40.4] Model magnitude in r filter AB scale (r) (7) \\
imag & Float & mag & [2.6/39.5] Model magnitude in i filter AB scale (i) (7) \\
zmag & Float & mag & [3.5/37.2] Model magnitude in z filter AB scale (z) (7) \\
e\_umag & Double & mag & [8.9e-8/3444] Mean error on umag (err\_u) \\
e\_gmag & Double & mag & [3.8e-8/5955] Mean error on gmag (err\_g) \\
e\_rmag & Double & mag & [5.2e-8/915] Mean error on rmag (err\_r) \\
e\_imag & Double & mag & [0/634] Mean error on imag (err\_i) \\
e\_zmag & Double & mag & [9.4e-8/3529] Mean error on zmag (err\_z) \\
zsp & Double &  & [-0.012/7.06] Spectroscopic final redshift (when SpObjID>0) (spec\_z) (6) \\
e\_zsp & Double &  & [-6/30689] Mean error on zsp (negative for bad fit) (spec\_zErr) (6) \\
f\_zsp & Short &  & [0/902] Zwarning flag (spec\_zWarning) (6)(9) \\
zph & Float &  & [5e-4/1] Photometric redshift estimated by robust fit to nearest neighbors \\
 & & & in a reference set (photoz\_z) (12) \\
e\_zph & Float &  & [0.003/0.3] Estimated error of the photometric redshift (photoz\_zErr) (12) \\
$\langle \mathrm{zph}\rangle$ & Float &  & [0.009/0.9] Average redshift of the nearest neighbors; if significantly different \\
 & & & from zph this might be a better estimate than zph (photoz\_nnAvgZ) (12) \\
Q & Short &  & [1/3] Quality of the observation: \\
 & & & 1$=$bad; 2$=$barely acceptable; 3$=$good (field\_quality) (13) \\
SDSS16 & String &  & SDSS-DR16 name based on J2000 position; provided by CDS \\
Sp-ID & String &  & Spectroscopic Plate-MJD-Fiber identifier (plate-mjd-fiberID) \\
\hline
\multicolumn{4}{c}{Columns for Pan-STARRS DR1 \citep{chambers_pan-starrs1_2016}} \\
RAJ2000 & Double & deg & Right ascension (J2000) (weighted mean) at epoch "Epoch" (raMean) (1) \\
DEJ2000 & Double & deg & Declination (J2000) (weighted mean) at epoch "Epoch" (decMean) (1) \\
objID & Long &  & Unique object identifier (objID) \\
f\_objID & Long &  & Information flag bitmask indicating details of the photometry (objInfoFlag) (2) \\
Qual & Short &  & Subset of objInfoFlag denoting whether this object is real \\
 & & & or a likely false positive (qualityFlag) (3) \\
e\_RAJ2000 & Double & arcsec &  Right ascension standard deviation from single epoch detections (raMeanErr) \\
e\_DEJ2000 & Double & arcsec &  Declination standard deviation from single epoch detections (decMeanErr) \\
Epoch & Double & d &  Modified Julian Date (MJD) of the mean epoch (epochMean) \\
Ns & Short &  & [0/50] Number of stack detections (nStackDetections) \\
Nd & Short &  & [3/369] Number of single epoch detections in all filters (nDetections) \\
gmag & Double & mag & [-33/33] Mean PSF AB magnitude from g filter (4866{AA}) detections (gMeanPSFMag) \\
e\_gmag & Float & mag & [0/0.5] Error in gmag (gMeanPSFMagErr) \\
gKmag & Double & mag &  Mean Kron (1980) AB magnitude from g filter detections (gMeanKronMag) \\
e\_gKmag & Float & mag &  Error in gKmag (gMeanKronMagErr) \\
gFlags & Integer &  & Information flag bitmask for mean object from g filter detections (gFlags) (4) \\
rmag & Double & mag & [-6/30] Mean PSF AB magnitude from r filter (6215{AA}) detections (rMeanPSFMag) \\
e\_rmag & Float & mag & [0/0.5] Error in rmag (rMeanPSFMagErr) \\
rKmag & Double & mag &  Mean Kron (1980) AB magnitude from r filter detections (rMeanKronMag) \\
e\_rKmag & Float & mag &  Error in rKmag (rMeanKronMagErr) \\
rFlags & Integer &  & Information flag bitmask for mean object from r filter detections (rFlags) (4) \\
imag & Double & mag & [-5/30] Mean PSF AB magnitude from i filter (7545{AA}) detections (iMeanPSFMag) \\
e\_imag & Float & mag & [0/0.5] Error in imag (iMeanPSFMagErr) \\
iKmag & Double & mag &  Mean Kron (1980) AB magnitude from i filter detections (iMeanKronMag) \\
e\_iKmag & Float & mag &  Error in iKmag (iMeanKronMagErr) \\
iFlags & Integer &  & Information flag bitmask for mean object from i filter detections (iFlags) (4) \\
zmag & Double & mag & [-5/29] Mean PSF AB magnitude from z filter (8679{AA}) detections (zMeanPSFMag) \\
e\_zmag & Float & mag & [0/0.5] Error in zmag (zMeanPSFMagErr) \\
zKmag & Double & mag &  Mean Kron (1980) AB magnitude from z filter detections (zMeanKronMag) \\
e\_zKmag & Float & mag &  Error in zKmag (zMeanKronMagErr) \\
zFlags & Integer &  & Information flag bitmask for mean object from z filter detections (zFlags) (4) \\
ymag & Double & mag & [-5/29] Mean PSF AB magnitude from y filter (9633{AA}) detections (yMeanPSFMag) \\
e\_ymag & Float & mag & [0/0.5] Error in ymag (yMeanPSFMagErr) \\
yKmag & Double & mag &  Mean Kron (1980) AB magnitude from y filter detections (yMeanKronMag) \\
e\_yKmag & Float & mag &  Error in yKmag (yMeanKronMagErr) \\
yFlags & Integer &  & Information flag bitmask for mean object from y filter detections (yFlags) (4) \\
\hline
\multicolumn{4}{c}{Columns for AllWISE \citep{cutri_vizier_2021}} \\
AllWISE & String &  & WISE All-Sky Release Catalog name, based on J2000 position, (designation) \\
RAJ2000 & Double & deg & Right ascension (J2000) \\
DEJ2000 & Double & deg & Declination (J2000) \\
Im & String &  & Image of the 4 bands from the IRSA server (IPAC) \\
w1mpro & Float & mag &  W1 magnitude (3.35um) \\
w1sigmpro & Float & mag &  Mean error on W1 magnitude \\
w2mpro & Float & mag &  W2 magnitude (4.6um) \\
w2sigmpro & Float & mag &  Mean error on W2 magnitude \\
w3mpro & Float & mag &  W3 magnitude (11.6um) \\
w3sigmpro & Float & mag &  Mean error on W3 magnitude \\
w4mpro & Float & mag &  W4 magnitude (22.1um) \\
w4sigmpro & Float & mag &  Mean error on W4 magnitude \\
Jmag & Float & mag &  2MASS J magnitude (1.25um) \\
e\_Jmag & Float & mag &  Mean error on J magnitude \\
Hmag & Float & mag &  2MASS H magnitude (1.65um) \\
e\_Hmag & Float & mag &  Mean error on H magnitude \\
Kmag & Float & mag &  2MASS Ks magnitude (2.17um) \\
e\_Kmag & Float & mag &  Mean error on Ks magnitude \\
cc\_flags & String &  & [0DHOPdhop] Contamination and confusion flag, one per band (cc\_flags) (2) \\
ext\_flg & Short &  & [0-5] Extended source flag (ext\_flg) (3) \\
var\_flg & String &  & Variability flag, one per band (var\_flg) (4) \\
ph\_qual & String &  & [ABCUXZ] Photometric quality flag (8) \\
\hline
\multicolumn{4}{c}{Columns for unWISE \citep{schlafly_unwise_2019}} \\
unwise\_objid & char &  & unique object ID \\
ra & double & deg & right ascension: W1 position, if available; otherwise W2 position \\
dec & double & deg & declination: W1 position, if available; otherwise W2 position \\
flux\_1 & double & 'nMgy' & flux, band 1 \\
dflux\_1 & double & 'nMgy' & statistical uncertainty in flux\_1 \\
flux\_2 & double & 'nMgy' & flux, band 2 \\
dflux\_2 & double & 'nMgy' & statistical uncertainty in flux\_2 \\
fluxlbs\_1 & double & 'nMgy' & local-background-subtracted flux, band 1 \\
dfluxlbs\_1 & double & 'nMgy' & uncertainty in fluxlbs\_1 \\
fluxlbs\_2 & double & 'nMgy' & local-background-subtracted flux, band 2 \\
dfluxlbs\_2 & double & 'nMgy' & uncertainty in fluxlbs\_2 \\
qf\_1 & double &  & quality factor, band 1: PSF-weighted fraction of pixels contributing to this detection \\
qf\_2 & double &  & quality factor, band 2: PSF-weighted fraction of pixels contributing to this detection \\
rchi2\_1 & double &  & average $\mathrm{\chi^2}$ per pixel, weighted by PSF, band 1 \\
rchi2\_2 & double &  & average $\mathrm{\chi^2}$ per pixel, weighted by PSF, band 2 \\
fracflux\_1 & double &  & fraction of flux in this object's PSF that comes from this object, band 1 \\
fracflux\_2 & double &  & fraction of flux in this object's PSF that comes from this object, band 2 \\
spread\_model\_1 & double &  & SExtractor spread\_model parameter, band 1 \\
dspread\_model\_1 & double &  & uncertainty in spread\_model\_1 \\
spread\_model\_2 & double &  & SExtractor spread\_model parameter, band 2 \\
dspread\_model\_2 & double &  & uncertainty in spread\_model\_2 \\
fwhm\_1 & double & pix & full-width at half-maximum of PSF, band 1 \\
fwhm\_2 & double & pix & full-width at half-maximum of PSF, band 2 \\
\hline
\multicolumn{4}{c}{Columns for FIRST \citep{helfand_last_2015}} \\
FIRST & String &  & FIRST Source designation \\
RAJ2000 & String &  & Right Ascension J2000 \\
DEJ2000 & String &  & Declination J2000 \\
p(S) & Float &  & [0,1] Probability of being a side lobe \\
Fpeak & Double & mJy & Peak flux density at 1.4GHz \\
Fint & Double & mJy & Integrated flux density at 1.4GHz \\
Rms & Float & mJy & Local noise estimate \\
Maj & Float & arcsec & Major axis (FWHM) \\
Min & Float & arcsec & Minor axis (FWHM) \\
PA & Float & deg & [0/180] Position angle \\
N1 & String &  & [0/10] Number of SDSS-DR10 counterparts \\
c1 & Character &  & [sg-] SDSS class: s=star, g=galaxy \\
N2 & String &  & Number of 2MASS counterparts \\
\hline
\multicolumn{4}{c}{Columns for LoTSS DR2 \citep{shimwell_lofar_2022}} \\
Source & String &  & The radio name of the source, automatically generated from RA and DEC \\
RAJ2000 & Double & deg & Right ascension (J2000) (RA) \\
DEJ2000 & Double & deg & Declination (J2000) (DEC)  \\
Speak & Double & mJy/beam & The peak Stokes I flux density per beam of the source (Peak\_flux) \\
SpeakTot & Double & mJy & The total, integrated Stokes I flux density of the source \\
 & & & at the reference frequency (Total\_flux) \\
Maj & Float & arcsec & FWHM of the major axis of the source, INCLUDING convolution \\ 
 & & & with the 6-arcsec LOFAR beam (Maj) \\
Min & Float & arcsec & FWHM of the minor axis of the source, INCLUDING convolution \\  & & & with the 6-arcsec LOFAR beam (Min) \\
DCMaj & Float & arcsec & The FWHM of the major axis of the source, after de-convolution \\
 & & & with the 6-arcsec LOFAR beam (DC\_Maj) \\
DCMin & Float & arcsec & The FWHM of the minor axis of the source, after de-convolution \\
 & & & with the 6-arcsec LOFAR beam (DC\_Min) \\
PA & Float & deg & The position angle of the major axis of the source measured east of north, \\
 & & & after de-convolution with the 6-arcsec LOFAR beam (PA) \\
DCPA & Float & deg & The position angle of the major axis of the source measured east of north, \\
 & & & after de-convolution with the 6-arcsec LOFAR beam (DC\_PA) \\
SCode & String &  & A code that defines the source structure \\
 & & & in terms of the fitted Gaussian components \\
MaskFract & Float &  & The fraction of the source that is in the CLEAN mask (Masked\_Fraction) \\
\hline
\multicolumn{4}{c}{Columns calculated in this work} \\
flag\_multi\_opt  & int &  & 0=VLASS source has unique UNIONS counterpart within the search radius; \\
&  &  & 1=VLASS source has multiple UNIONS counterparts within the search radius; \\
sep\_guc\_vlass & float & arcsec & The angular separation between UNIONS and VLASS coordinates \\
sep\_guc\_first & float & arcsec & The angular separation between UNIONS and FIRST coordinates \\
sep\_guc\_lotss & float & arcsec & The angular separation between UNIONS and LoTSS coordinates \\
sep\_vlass\_first & float & arcsec & The angular separation between VLASS and FIRST coordinates \\
sep\_vlass\_lotss & float & arcsec & The angular separation between VLASS and LoTSS coordinates \\
sep\_first\_lotss & float & arcsec & The angular separation between FIRST and LoTSS coordinates \\
alpha\_144\_1400 & float &  & The spectral index between LoTSS 144~MHz and FIRST 1.4~GHz \\
alpha\_144\_1400\_err & float &  & Uncertainties of alpha\_144\_1400 \\
alpha\_1400\_3000 & float &  & The spectral index between FIRST 1.4~GHz and VLASS 3~GHz \\
alpha\_1400\_3000\_err & float &  & Uncertainties of alpha\_1400\_3000 \\
alpha\_144\_3000 & float &  & The spectral index between LoTSS 144~MHz and VLASS 3~GHz \\
alpha\_144\_3000\_err & float &  & Uncertainties of alpha\_144\_3000 \\
r\_obs & float &  & Observed radio loudness defined as the ratio of 1.4~GHz to $i$-band flux densities \\
luminosity\_1p4 & float & W/Hz & Spectral luminosity at 1.4~GHz, VLASS 3~GHz scaled if no FIRST detection \\
luminosity\_3p0 & float & W/Hz & Spectral luminosity at 3~GHz, alpha=-0.7 used if no FIRST detection \\
mag\_vlass & float & mag & Radio AB magnitude for VLASS \\
mag\_first & float & mag & Radio AB magnitude for FIRST \\
ls\_vlass & float & kpc & VLASS projected linear size calculated from the deconvolved major axis \\
ls\_lotss & float & kpc & LoTSS projected linear size calculated from the deconvolved major axis \\
flag\_z\_photo & int &  & -1$=$no valid photo-$z$; \\
& & & 0$=$valid $\mathrm{UNIONS5000}_{ugriz}$ photo-$z$; 1$=$valid $\mathrm{UNIONS5000}_{ugri}$ photo-$z$\\
Z\_B\_ref & float &  & if flag\_z\_photo=-1, $\mathrm{UNIONS5000}_{ugriz}$ photo-$z$ used but not reliable; \\
& & & if flag\_z\_photo=0, $\mathrm{UNIONS5000}_{ugriz}$ photo-$z$ used; \\
& & & if flag\_z\_photo=1, $\mathrm{UNIONS5000}_{ugri}$ photo-$z$ used;\\
Z\_ref & float &  & preferred redshifts; Z\_ref = SDSS spec-$z$ if available; otherwise Z\_ref = Z\_B\_ref\\
extended & int &  & 1=Extended; 0=Not Extended. Classified following the criterion in \S\,\ref{subsec:extended sources} \\
\hline

\end{longtable}

\bibliography{references}{}
\bibliographystyle{aasjournal}

\end{CJK*}
\end{document}